\documentclass[5p,authoryear,10pt]{elsarticle}

\usepackage{amssymb}
\usepackage{amsmath}
\usepackage{natbib}
\usepackage{subcaption}
\usepackage{graphicx}
\usepackage{epstopdf}
\usepackage{marvosym}
\usepackage{epsfig}
\usepackage{multirow}
\usepackage{tikz}
\usepackage{nicefrac}
\usepackage{arydshln}
\usepackage{url}
\usepackage{blkarray}
\usepackage{hyphenat} 
\usepackage{lipsum}
\usepackage{caption}
\usepackage{booktabs}
\usetikzlibrary{arrows,calc,intersections,positioning,arrows,decorations.pathmorphing,decorations.markings,shapes}

\newcommand{\RN}[1]{
  \textup{\uppercase\expandafter{\romannumeral#1}}
}
\journal{Planetary and Space Science}

\begin{document}

\begin{frontmatter}

\title{Decoupled and coupled moons' ephemerides estimation strategies \\ Application to the JUICE mission}

\author[TUDELFT]{M. Fayolle \corref{cor}}
\ead{m.s.fayolle-chambe@tudelft.nl}

\author[TUDELFT]{D. Dirkx}
\ead{d.dirkx@tudelft.nl}
\author[IMCCE]{V. Lainey}
\ead{Valery.Lainey@imcce.fr}
\author[JIVE,TUDELFT]{L.I. Gurvits}
\ead{LGurvits@jive.eu}
\author[TUDELFT]{P.N.A.M. Visser}
\ead{P.N.A.M.Visser@tudelft.nl}

\address[TUDELFT]{Delft University of Technology, Kluyverweg 1, 2629HS Delft, The Netherlands}
\address[IMCCE]{IMCCE, Observatoire de Paris, 77 Av. Denfert-Rochereau, 75014, Paris, France}
\address[JIVE]{Joint Institute for VLBI ERIC, Oude Hoogeveensedijk 4, 7991PD Dwingeloo, The Netherlands}

\begin{abstract}
When reconstructing natural satellites' ephemerides from space missions' tracking data, the dynamics of the spacecraft and natural bodies are often solved for separately, in a decoupled manner. Alternatively, the ephemeris generation and spacecraft orbit determination can be performed concurrently. This method directly maps the available data set to the estimated parameters' covariances while fully accounting for all dynamical couplings. It thus provides a statistically consistent solution to the estimation problem, whereas this is not directly ensured with the decoupled strategy. For the Galilean moons in particular, the JUICE mission provides a unique, although challenging, opportunity for ephemerides improvement.
For such a dynamically coupled problem, choosing between the two state estimation strategies will be influential. 
This paper quantifies the Galilean moons' state uncertainties attainable when applying a coupled estimation strategy to simulated JUICE data, and discusses the challenges that remain to be addressed to achieve such a coupled solution from real observations. 
We first provide a detailed, explicit formulation for the coupled approach, which was still missing in the literature although already used in past studies. We then assessed the relative performances of the two ephemerides generation techniques for the JUICE test case. To this end, we used both decoupled and coupled models on simulated JUICE radiometric data. We compared the resulting covariances for the Galilean moons' states, and 
showed that the decoupled approach yields slightly lower formal errors for the moons' tangential positions. 
However, the coupled model can reduce the state uncertainties by more than one order of magnitude in the radial direction (\textit{i.e.} towards the central body). It also proved more sensitive to the dynamical coupling between Io, Europa and Ganymede, allowing the state solutions for the first two moons to fully benefit from JUICE orbital phase around Ganymede. \textcolor{black}{On the other hand, we showed that the choice of state estimation methods does not strongly affect the moons' gravity field determination.} 
Many issues still remain to be solved before a concurrent estimation strategy can be successfully applied, especially to reconstruct the moons' dynamics over long timescales. Nonetheless, our analysis highlights promising ephemerides improvements and thus motivates future efforts to reach a coupled state solution for the Galilean moons.
\end{abstract}


\end{frontmatter}

\section{Introduction} \label{sec:introduction}

\begin{figure}[b!]
	\centering
	\includegraphics[width=0.5\textwidth]{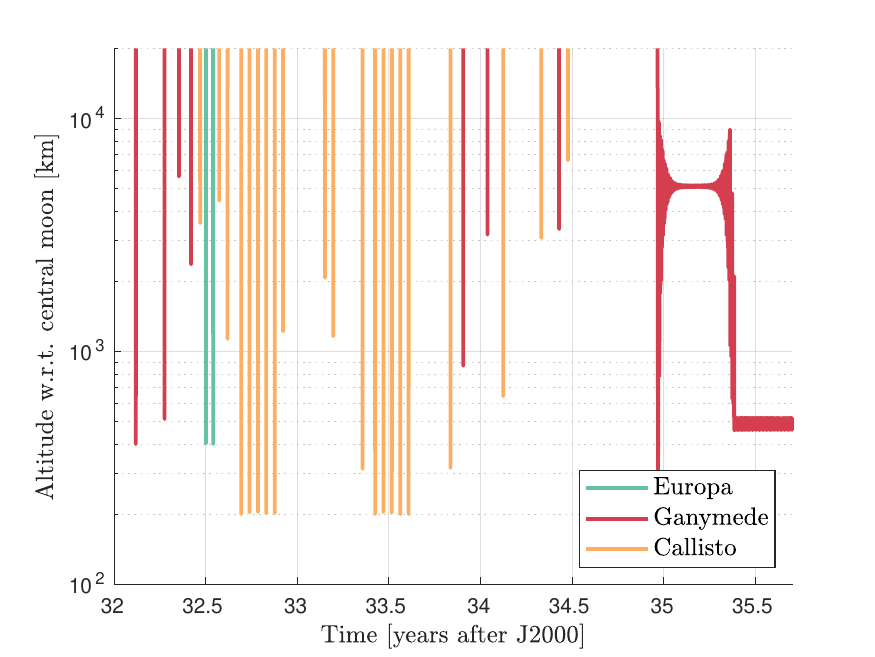}
	\caption{\textcolor{black}{Altitude of the JUICE spacecraft with respect to the Galilean moons during the flyby and orbital phases, based on CReMA 5.0.} The vertical lines directly provide the closest-approach distances for the flyby series, while the orbit of JUICE around Ganymede is clearly identifiable, starting slightly before 2035.}
	\label{fig:flybysAltitudes}
\end{figure}

The upcoming JUICE mission\footnote{https://sci.esa.int/web/juice} (JUpiter ICy moons Explorer) will focus on the three Galilean moons Europa, Ganymede and Callisto. The JUICE spacecraft is expected to arrive in the Jovian system in 2031, with a launch planned in 2023. It will first execute a series of flybys (2, 7 and 21 flybys at Europa, Ganymede and Callisto, respectively), from 2032 to 2034. \textcolor{black}{JUICE will then initiate its orbital phase around Ganymede, with an eccentricity ranging from 0.6 to 0 (GEO and GCO500 phases: Ganymede Elliptic Orbit and Ganymede Circular Orbit with an altitude of 5000 km, respectively).} \textcolor{black}{In May 2035, after a second elliptical phase, the spacecraft will eventually enter its final circular orbit at 500 km altitude (denoted GCO500), for a nominal period of 4 months.} This mission profile, displayed in Figure \ref{fig:flybysAltitudes} and adopted in the rest of this paper, was obtained from the version 5.0 of the CReMA (Consolidated Report on the Mission Analysis) \footnote{https://www.cosmos.esa.int/web/spice/spice-for-juice}.

\textcolor{black}{The JUICE spacecraft will carry one dedicated radio science instrument \citep[3GM: Gravity and Geophysics of Jupiter and the Galilean Moons, \textit{e.g.}][]{diBenedetto2021}, which will provide highly accurate range and Doppler measurements (see Section \ref{sec:estimationSettings}). These 3GM observations will be complemented by the PRIDE experiment \citep[Planetary Radio Interferometry and Doppler Experiment, \textit{e.g.}][]{gurvits2013}. The latter does not require any additional onboard instrument and uses tracking or 3GM radiometric signals to derive angular position measurements of the spacecraft with respect to the ICRF (International Celestial Reference Frame), as well as supplementary Doppler observables \citep[\textit{e.g.}][]{duev2012, bocanegra2018,molera2021}. The radiometric data to be acquired by both 3GM and PRIDE are expected to contribute to a more accurate determination of the Galilean moons' states \citep{dirkx2016,dirkx2017,lari2019,cappuccio2020}.} Improved ephemerides are crucial to better understand the long-term thermal-orbital evolution of these moons, which is strongly driven by tidal dissipation in both Jupiter and the satellites themselves \citep{peale1999,hussmannSpohn2004,greenberg2010,hay2020}. The moons' ephemerides provide a natural way to extract the current rates of tidal dissipation, through the observed migration rates of the satellites \citep[\textit{e.g.}][]{laineyTobie2005, lainey2009}. Furthermore, a better characterisation of tidal dissipation mechanisms can provide tighter constraints on the moons' interiors, which is critical to investigate sub-surface ocean's properties \citep[or confirm the existence of a putative ocean for Callisto, \textit{e.g.}][]{lunine2017}.

For natural satellites' ephemerides, the estimations of the spacecraft's and moons' dynamics are typically not performed in a coupled manner \citep[\textit{e.g.}][]{rosenblatt2008,durante2019}. Instead, when ephemerides are to be determined from flybys, the spacecraft trajectory with respect to the central body (\textit{i.e.} body at which the flyby is performed) is determined from the available tracking data, along with the central body's state at the flyby epoch. The per-flyby state solutions for the natural body define the so-called \textit{normal points}, which are then used in a second global estimation to reconstruct the long-term dynamics of this body \citep[\textit{e.g.}][]{durante2019}. 

If needed, a unified model may also be used, in which the spacecraft dynamics are determined in a multi-arc fashion, and the natural bodies' dynamics in a single-arc fashion, during a single inversion. Such an approach, used for instance by \cite{dirkx2018,lari2019}, has the advantage of automatically incorporating all dynamical couplings, as well as the full sensitivity to physical parameters of both types of dynamics. The decoupled strategy, on the other hand, reconstructs the moons' dynamics from the normal points, which only capture the moons' kinematics (and not dynamics) at each flyby. However, while desirable, a coupled solution for the spacecraft's and moons' states is not always achievable in practice \citep[\textit{e.g.}][]{durante2019}. It indeed requires the moons' dynamical models (with respect to the planet) to be consistent over both short and long timescales, to the accuracy level of the spacecraft's dynamics (with respect to the moon). \textcolor{black}{Here, short and long timescales respectively refer to typical flyby duration (\textit{i.e.} a few hours) and entire mission timeline (\textit{i.e.} several years, so still short with respect to system evolution's timescale)}.

For the JUICE mission, we will be presented with a unique situation: the mission profile \textcolor{black}{indeed} calls for a combination of flybys around multiple satellites and an extended orbit phase around Ganymede, which was never before performed in a planetary mission. \textcolor{black}{Additionally,} three of the four Galilean moons are in resonance, making the estimation of the different moons' dynamics strongly coupled, with the added complication that JUICE will provide a strong imbalance in data for these three moons. As a result, the estimation of ephemerides from JUICE-only data is (close to be) an ill-posed mathematical problem \citep{dirkx2017}. Due to the complexity and novelty of the mission profile, and to the strong dynamical couplings that are involved, the concurrent single- and multi-arc estimation strategy appears particularly well-suited for the JUICE test case. In this paper, we compare the simulated state estimation solutions obtained with both the decoupled and coupled approaches, to quantify the impact of the adopted estimation strategy. 

We limit ourselves to a covariance analysis, complemented by a deterministic simulation performed as verification. As already highlighted, the practical applicability of the coupled method to the JUICE mission is however not guaranteed, as bringing the dynamical models fidelity down to the required accuracy level will be really challenging. By definition, these issues cannot be addressed by a covariance analysis, as the resulting formal uncertainties do not account for inaccuracies in the dynamical models used for the moons and the JUICE spacecraft, or in the models representing the observations' errors.  Our paper thus assesses which uncertainty levels could be obtained with a coupled estimation, provided that our dynamical models are accurate enough for a viable solution to be achieved. The formal uncertainties we obtain therefore quantify the coupled strategy's requirements in terms of dynamical modelling accuracy. Besides comparison purposes, precisely evaluating the performance of the decoupled method is thus also crucial in case the obtention of a global coupled solution for the Galilean moons remains beyond (current) modelling capabilities. It must be noted that modelling issues, while not directly addressed by our covariance analysis, remain nonetheless deeply relevant for this study and will therefore be extensively discussed (see Sections \ref{sec:scopeComparativeAnalysis} and \ref{sec:modellingIssues}). More generally, the limitations and scope of our formal analysis will be further detailed in Section \ref{sec:scopeComparativeAnalysis}.

\textcolor{black}{The details of the coupled model, as well as the issues associated with its implementation and application, are not found in the literature}, despite its application in a number of past studies \citep[\textit{e.g.}][]{dirkx2018,lari2019,magnanini2021}. Therefore, we choose to provide a detailed exposition of our coupled method in Section \ref{sec:estimationFramework}, completed by a shorter description of the decoupled approach. The models used to either propagate the moons' and spacecraft's dynamics or simulate the JUICE radiometric observations are then described in Section \ref{sec:dynamicsAndObservations}. Section \ref{sec:results} presents the results of our comparative analysis of the coupled and decoupled estimation strategies, first for the flyby phase only, and then for the entire JUICE mission. Finally, Section \ref{sec:discussion} discusses in more detail the strengths and challenges of both estimation methods, before our conclusions are summarised in Section \ref{sec:conclusions}.

\section{Estimation framework} \label{sec:estimationFramework}

This section describes the whole estimation process, for both the coupled and decoupled approaches introduced in Section \ref{sec:introduction}. The complete formulation for the coupled single- and multi-arc state estimation model, still missing in the literature, is provided in Section \ref{sec:coupledModel}. For the sake of completeness, our implementation of the decoupled strategy for the JUICE case is given in Section \ref{sec:decoupledModel}. This section thus directly highlights the main differences between the two estimation strategies.

All methods described in the following were implemented in our TU Delft Astrodynamics Toolbox (Tudat) software\footnote{Documentation: \url{https://tudat-space.readthedocs.io}\\ Full source code: \url{https://github.com/tudat-team/tudat-bundle}}, and are therefore freely available. 

\subsection{Covariance analysis} \label{sec:covarianceAnalysis}

We first briefly review the propagation of the variational equations and describe how covariance matrices are generated and propagated in our simulations, as typically implemented in any estimation process, either single- and/or multi-arc. 

\subsubsection{Variational equations formulation}
\label{sec:variationalEquations}

The variational equations describe how the dynamics of the system are influenced by the parameters to be estimated. In the following, we adopt the nomenclature of \cite{montenbruckGill2000}. The state vector is denoted as $\mathbf{y}$ and is propagated numerically from the initial time $t_{0}$ using
\begin{align}
\dot{\mathbf{y}}(t)=\mathbf{f}(\mathbf{y},\mathbf{p},t),\label{eq:basicStateDifferential}
\end{align}
where $\mathbf{p}$ is a vector of parameters influencing the system's dynamics \textcolor{black}{or the observations,} and $\mathbf{f}$ represents the dynamical model (described in Section \ref{sec:dynamicsAndObservations}). Unless otherwise indicated, all states are expressed in a reference frame with inertial orientation (\emph{e.g.} J2000).

We stress that, in a general formulation, the states $\mathbf{y}$ need not be translational states, but may be any type of dynamics, of any number of bodies (see \cite{mazarico2017,dirkx2019} for an example of coupled translational-rotational dynamics estimation of multiple bodies).

The state transition matrix $\boldsymbol{\Phi}(t,t_{0})$ and sensitivity matrix $\mathbf{S}(t)$ are defined as
\begin{align}
\boldsymbol{\Phi}(t,t_{0})&=\frac{\partial\mathbf{y}(t)}{\partial\mathbf{y}(t_{0})},\\
\mathbf{S}(t)&=\frac{\partial\mathbf{y}(t)}{\partial\mathbf{p}}.
\end{align}
The differential equations used to solve for $\boldsymbol{\Phi}$ and $\mathbf{S}$ are termed the variational equations, and are given by
\textcolor{black}{\begin{align}
	\frac{d\boldsymbol{\Phi}(t,t_{0})}{dt}&=\frac{\partial\mathbf{f}(\mathbf{y},\mathbf{p},t)}{\partial\mathbf{y}(t)}\boldsymbol{\Phi}(t,t_{0}),\label{eq:basicVariationalPhi}\\
	\frac{d\mathbf{S}(t)}{dt}&=\frac{\partial\mathbf{f}(\mathbf{y},\mathbf{p},t)}{\partial\mathbf{y}(t)}\mathbf{S}(t)+\frac{\partial\mathbf{f}(\mathbf{y},\mathbf{p},t)}{\partial\mathbf{p}},\label{eq:basicVariationalS}
	\end{align}}
with the following initial conditions:
\begin{align}
\boldsymbol{\Phi}(t_{0},t_{0})&=\mathbf{1}_{n\times n}, \label{eq:initialConditionPhi}\\
\mathbf{S}(t_{0})&=\mathbf{0}_{n\times n_{p}}, \label{eq:initialConditions}
\end{align}
where $n$ and $n_{p}$ represent the sizes of the state vector $\mathbf{y}$ and parameter vector $\mathbf{p}$, respectively. The single-arc and multi-arc formulations are essentially identical, with the sole difference that the multi-arc solution is obtained by subsequent, independent, integrations of Eqs. (\ref{eq:basicStateDifferential}), (\ref{eq:basicVariationalPhi}) and (\ref{eq:basicVariationalS}).

A variant of the multi-arc method, referred to as the constrained multi-arc approach, uses the fact that the arc-wise state estimates obtained for a given body should be consistent to further constrain the estimation solution \citep{alessi2012,serra2018}. In our analysis, we however chose to limit ourselves to the unconstrained multi-arc estimation. During the first part of the mission, the flybys are indeed temporarily distant, such that propagating information from previous arcs would not efficiently constrain the JUICE spacecraft's state. When arcs are contiguous (\textit{i.e.} orbital phase around Ganymede), the high quality of the estimation solution anyway undermines the use of multi-arc constraints.

\subsubsection{Propagated covariance}
\label{sec:covariancePropagation}

Let \textcolor{black}{$\mathbf{h}(T,\mathbf{q})$} denote the set of all modelled observations generated up to a time $T$. The design matrix \textcolor{black}{$\mathbf{H}(T,\mathbf{q})$} associated with these observations is then formed by computing 
\begin{align}
\mathbf{H}(T,\mathbf{q})&=\frac{\partial \mathbf{h}(T,\mathbf{q})}{\partial \mathbf{q}},\label{eq:partialsMatrix}
\end{align}
with $\mathbf{q}$ a vector containing the estimated parameters \citep[\textit{e.g.}][]{montenbruckGill2000, milaniGronchi2010}. \textcolor{black}{It usually includes initial states parameters, represented by the vector $\mathbf{y}_0$, and a subset of the parameters $\mathbf{p}$ influencing the dynamical or observational models. To simplify the notations, the vector of estimated parameters $\mathbf{q}$ will be divided as $\mathbf{q}=[\mathbf{y}_{0};\mathbf{p}]$ in the following. It should however be noted that the exact definition of $\mathbf{q}$ depends on the estimation model, and that $\mathbf{y}_0$ and $\mathbf{p}$ might not directly incorporate all initial states, dynamical and observational models. More details on how $\mathbf{y}_0$ and $\mathbf{p}$ are precisely defined for both the coupled and decoupled models will be provided in Sections \ref{sec:coupledModel}, \ref{sec:decoupledModel} and  \ref{sec:estimationSettings}.}

The covariance matrix of $\mathbf{q}$ obtained using data up to time $T$ is denoted $\mathbf{P}_{\mathbf{q}\mathbf{q}}(T)$ and is given by
\begin{align}
\mathbf{P}_{\mathbf{q}\mathbf{q}}(T)&=\left(\mathbf{P}_{\mathbf{q}\mathbf{q},0}^{-1}+\left(\mathbf{H}^{T}(T)\mathbf{W}(T)\mathbf{H}(T) \right)\right)^{-1},\label{eq:covariance}
\end{align}
where $\mathbf{P}_{\mathbf{q}\mathbf{q},0}$ is the \emph{a priori} covariance matrix of the parameters $\mathbf{q}$ (see Section \ref{sec:aPrioriKnowledge} for \textit{a priori} knowledge in our JUICE test case). \textcolor{black}{The matrix $\mathbf{W}(T)$ contains the weights associated with all observations up to time $T$. In most cases, it is set as a diagonal matrix with $W_{ii}=\sigma_{h,i}^{-2}$, implicitly assuming the measurement uncertainties to be uncorrelated. This is however not the case in every estimation step of the decoupled model, as will be discussed in Sections \ref{sec:generalPrincipleDecoupled} and \ref{sec:aPrioriKnowledge}.} $\sigma_{h,i}$ denotes the uncertainty of observation $i$. The covariance $\mathbf{P}_{\mathbf{q}\mathbf{q}}(T)$ can be used to compute the covariance of the state $\mathbf{y}$ at any later time $t$. We refer to this propagated covariance as $\mathbf{P}_{\mathbf{y}\mathbf{y}}(t,T)$ and define it as
\begin{align}
\mathbf{P}_{\mathbf{y}\mathbf{y}}(t,T)=[\boldsymbol{\Phi}(t,t_{0});\mathbf{S}(t)]\mathbf{P}_{\mathbf{q}\mathbf{q}}(T)[\boldsymbol{\Phi}(t,t_{0});\mathbf{S}(t)]^{T},\label{eq:propagatedCovariance}
\end{align}
where $\mathbf{\Phi}$ and $\mathbf{S}$ are the state transition and sensitivity matrices obtained through Eqs. \ref{eq:basicVariationalPhi}, \ref{eq:basicVariationalS}, \ref{eq:initialConditionPhi} and \ref{eq:initialConditions}.
For the covariances in Eqs. (\ref{eq:covariance}) and (\ref{eq:propagatedCovariance}), the formal errors are obtained from the square root of the diagonal elements of \textcolor{black}{$\mathbf{P_{qq}}$ and $\mathbf{P_{yy}}$, respectively}.

\subsection{Coupled single- and multi-arc estimation}
\label{sec:coupledModel}

This section describes the extension of the variational equations introduced in Section \ref{sec:variationalEquations} to the concurrent estimation of single- and multi-arc states. The formulation specifics are detailed in Section \ref{sec:coupledMethodJuice} for the JUICE case.

\subsubsection{General principle} \label{sec:generalPrincipleCoupled}

\begin{figure} [tbp!] 
	\centering
	\begin{minipage}[l]{0.45\textwidth}
		\centering
		\subcaptionbox{Coupled spacecraft and moons state estimation. The arc-wise spacecraft states and single-arc moon state are estimated concurrently in a single step. 
			\label{fig:coupledModel}}
		{\includegraphics[width=1.0\textwidth]{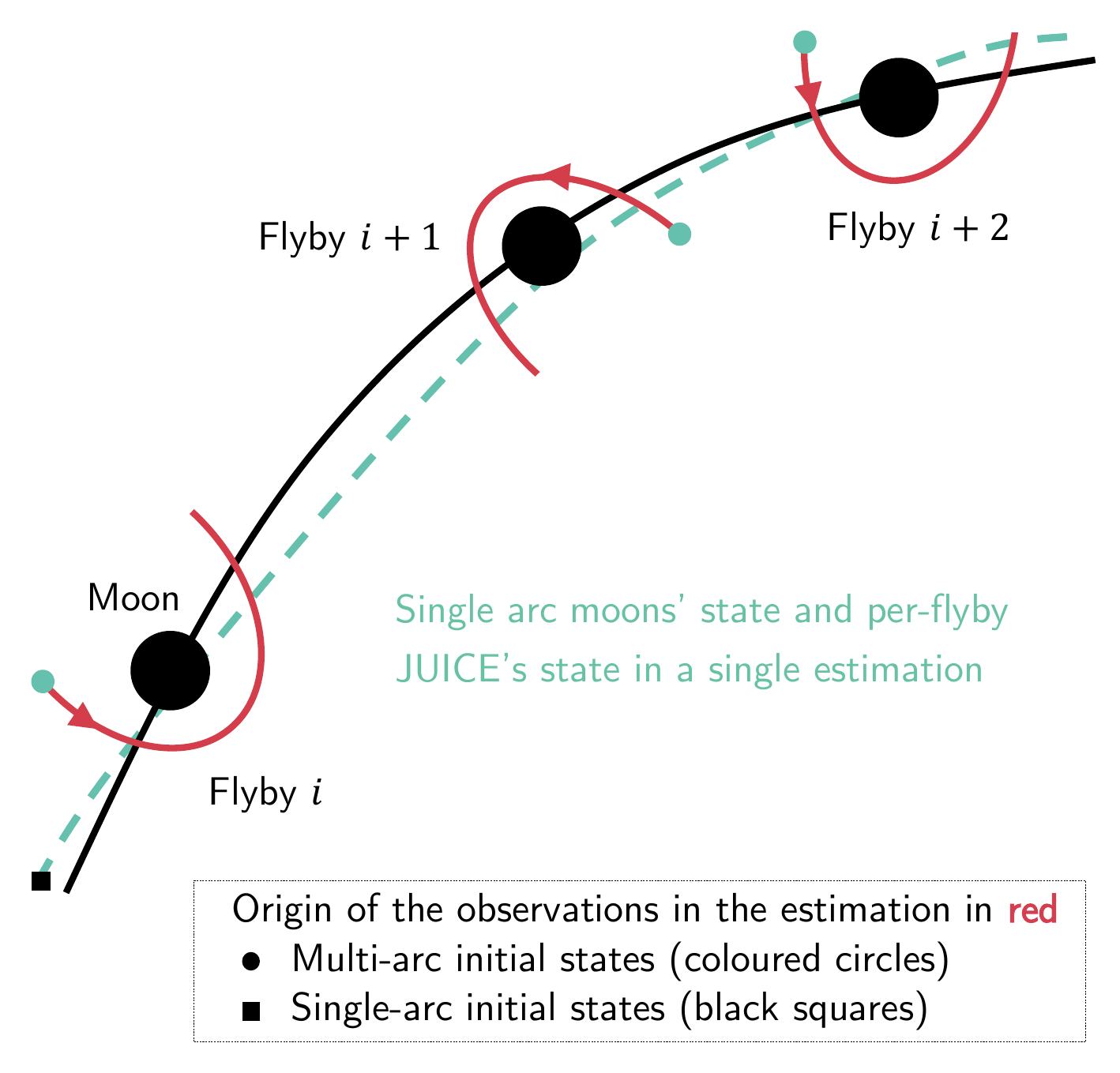}}
	\end{minipage}
	\hfill{} 
	\begin{minipage}[l]{0.45\textwidth}
		\centering
		\subcaptionbox{Decoupled spacecraft and moons state estimation. The arc-wise state solutions for the central moon, obtained at the end of the first step, are used as inputs for the second estimation step.
			\label{fig:decoupledModel}}
		{\includegraphics[width=1.0\textwidth]{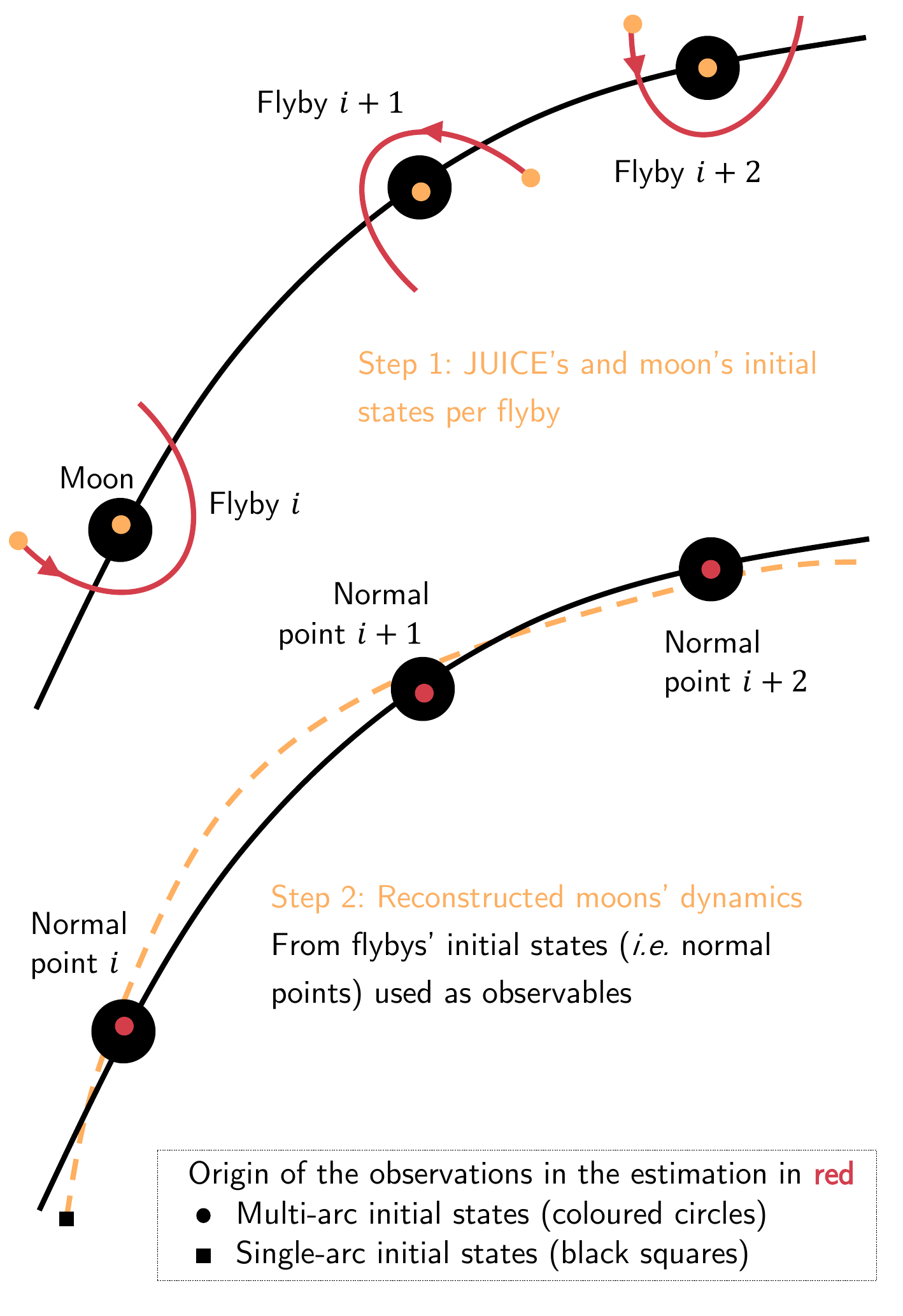}}
	\end{minipage}
	\caption{Schematic representations of the decoupled and coupled state estimations, specifically displayed for a series of JUICE flybys around a single moon. The solid lines represent the actual trajectories of either the spacecraft or the central moon, while the dashed lines depict the reconstructed moons' dynamics after the estimation process is complete.}
	\label{fig:models}
\end{figure}

The coupled strategy relies on the concurrent estimation of the spacecraft orbit and natural bodies' ephemerides, as well as of all parameters influencing the \textcolor{black}{dynamics and/or observations} (vector $\mathbf{q}$ in Section \ref{sec:covarianceAnalysis}). The natural bodies' dynamics, on the other hand, are reconstructed over a single arc. More precisely, the spacecraft orbit is solved for in an arc-wise manner (along with any observations- or spacecraft-related parameters, \textit{e.g.} biases, accelerometer calibrations factors, \textit{etc.}). Such a coupled model allows us to directly and robustly link observation strategies, data quality, mission profile, \textit{etc.} to the final formal uncertainties in natural bodies' ephemerides and dynamical parameters (tidal dissipation, gravity field coefficients). Figure \ref{fig:coupledModel} provides a schematic visualisation of this coupled approach, taking the JUICE flyby phase as an example. 

A preliminary framework for a concurrent single- and multi-arc estimation was briefly described by \cite{dirkx2018}. A similar method was also used by \cite{lari2019} to study the onset of chaos in the dynamics of  the JUICE spacecraft. The following sections provide the detailed and complete formulation for such a coupled estimation procedure for single- and multi-arc dynamics. 

\subsubsection{Coupled variational equations} \label{sec:coupledVariationalEquations}

We denote the single-arc state vector $\mathbf{y}_{_{S}}(t)$, of size $n_{s}$ and associated with the initial time $t_{0}$. The multi-arc state, on the other hand, is designated by $\mathbf{y}_{_{M}}(t)$ and the arc-wise initial time is noted $t_{i}$ for arc $i$, with $i=1...N$ for $N$ arcs. The size of each arc-wise state vector is $n_{{m}}$. 

The multi-arc state function $\mathbf{y}_{_{M}}(t)$ is defined as
\begin{align}
\mathbf{y}_{_{M}}(t)=\mathbf{y}_{_{M,i}}(t),\label{eq:multiArcStateFunction}\\
t\in[t_{i},\tilde{t}_{i}],\label{eq:arcTimeInterval}
\end{align}
where $\mathbf{y}_{_{M,i}}(t)$ refers to the state at time $t$ during arc $i$, and $\tilde{t}_{i}$ denotes the end time of arc $i$. It should be noted that the multi-arcs need not be contiguous, and gaps may exist in the arc-wise solutions to Eq. (\ref{eq:basicStateDifferential}). Eq. (\ref{eq:multiArcStateFunction}) is therefore only defined if an arc $i$ exists that satisfies Eq. (\ref{eq:arcTimeInterval}) at the time $t$.

The full state function is given as a combination of the single- and multi-arc states at time $t$, as follows:
\begin{align}
\mathbf{y}(t)=\begin{pmatrix}\mathbf{y}_{_{S}}(t)\\\mathbf{y}_{_{M}}(t)\end{pmatrix}.\label{eq:combinedStateVector}
\end{align}
For our estimation, we require a linearised model for the change in $\mathbf{y}(t)$ induced by a variation in the parameters $\mathbf{p}$ and in the full vector of initial states $\mathbf{y}_{0}$, to compute $\mathbf{S}$ and $\mathbf{\Phi}$, respectively:
\begin{align}
\mathbf{y}_{0}=\begin{pmatrix}\mathbf{y}_{_{S}}(t_{0})\\\mathbf{y}_{_{M,1}}(t_{1})\\\vdots\\\mathbf{y}_{_{M,N}}(t_{N})\end{pmatrix}. \label{eq:initialState}
\end{align}
\textcolor{black}{Looking at Eqs. (\ref{eq:combinedStateVector}) and (\ref{eq:initialState}), the size of the full initial state vector to be estimated ($\mathbf{y}_0$) is different from that of the state function $\mathbf{y}(t)$, as the former combines all single-arc and multi-arc initial states, while $\mathbf{y}(t)$ only includes the single-arc states and the multi-arc state of the current arc.} We note that $\mathbf{p}$ may affect the single- or multi-arc dynamics solutions, or both. The only limitation imposed on the dynamics (Eq. \ref{eq:basicStateDifferential}) is that the differential equation for $\mathbf{y}_{_{S}}(t)$ must be independent of $\mathbf{y}_{_{M}}(t)$. The opposite is not true, $\mathbf{y}_{_{M}}(t)$ being allowed to (and in our case does) depend on $\mathbf{y}_{_{S}}(t)$. These assumptions only hold if the masses of the multi-arc bodies are negligible with respect to the single-arc bodies'. This is typically the case as the spacecraft's dynamics are generally propagated in a multi-arc manner, while the bodies included in the single-arc solution are often natural bodies (see Section \ref{sec:coupledMethodJuice}).

We use $\boldsymbol{\Phi}_{_{SS}}(t,t_{0}), \boldsymbol{\Phi}_{_{MM,i}}(t,t_{i})$ to refer to the single- and multi-arc state transition matrices, respectively. Similarly, the single- and multi-arc sensitivity matrices are denoted as $\mathbf{S}_{_{S}}(t)$ and $\mathbf{S}_{_{M,i}}(t)$. We note that, for the multi-arc case, the parameter vector $\mathbf{p}$ can include local parameters that influence the dynamics of a single arc $i$ only, as well as global parameters that affect all arcs.

The state transition matrix is defined as the derivative of the current state $\mathbf{y}(t)$ (Eq. \ref{eq:combinedStateVector}) with respect to both the single-arc initial state $\mathbf{y}_{_{S}}(t_{0})$ at time $t_0$, and the arc-wise initial states $\mathbf{y}_{_{M,i}}$, at the beginning $t_i$ of each arc. The full state transition matrix, noted $\boldsymbol{\Phi}(t;t_{0},t_{i})$, and sensitivity matrix $\mathbf{S}(t)$ can thus be written as
\begin{align}
&\boldsymbol{\Phi}(t;t_{0},t_{i})=\frac{\partial\mathbf{y}(t)}{\partial\mathbf{y}_{0}} \label{eq:totalPhi} \\
&=\begin{pmatrix} \boldsymbol{\Phi}_{_{SS}}(t,t_{0}) & \mathbf{0}_{{n_{s},n_{m}(i-1)}} & \mathbf{0}_{{n_s,n_m}} &\mathbf{0}_{{n_s,n_m(N-i)}}
\\
\boldsymbol{\Phi}_{_{MS,i}}(t,t_{i})  & \mathbf{0}_{{n_m,n_m(i-1)}}& \boldsymbol{\Phi}_{_{MM}}(t,t_{i}) &\mathbf{0}_{{n_m,n_m(N-i)}}
\end{pmatrix}, \nonumber\\
&\mathbf{S}(t)=\frac{\partial\mathbf{y}(t)}{\partial\mathbf{p}}
=\begin{pmatrix}\mathbf{S}_{_{S}}(t)
\\\mathbf{S}_{_{M,i}}(t)
\end{pmatrix},\label{eq:totalS}
\end{align}
where we introduced the coupling term
\begin{align}
\boldsymbol{\Phi}_{_{MS,i}}(t,t_{i}) =\frac{\partial\mathbf{y}_{_{M,i}}(t)}{\partial\mathbf{y}_{_{S}}(t_{0})} \label{eq:coupledSTM}
\end{align}
into the state transition matrix. The zero entries in the first row of Eq. (\ref{eq:totalPhi}) directly result from the dynamics of $\mathbf{y}_{_{S}}(t)$ being independent of $\mathbf{y}_{_{M}}(t)$. For clarification purposes, the dimensions of these zero blocks are specified as subscripts.

To obtain the numerical solution to the coupled variational equations, we first propagate the single-arc dynamics and variational equations, to obtain $\mathbf{y}_{_{S}}$, $\mathbf{\Phi}_{_{SS}}$ and $\mathbf{S}_{_{S}}$. For the multi-arc formulation, the differential equations for $\boldsymbol{\Phi_{_{MM}}}$ are unchanged compared to the classical (decoupled) approach. However, to compute the full state transition and sensitivity matrices, we need a formulation for the coupling term $\mathbf{\Phi}_{_{MS,i}}$ which incorporates the influence of single-arc dynamics on the multi-arc dynamics, as follows:
\begin{align}
&\frac{d\boldsymbol{\Phi}_{_{MS,i}}(t,t_{i})}{dt}=\frac{\partial\dot{\mathbf{y}}_{_{M,i}}(t)}{\partial\mathbf{y}_{_{S}}(t)}\boldsymbol{\Phi}_{_{SS}}(t,t_{0})+\frac{\partial\dot{\mathbf{y}}_{_{M,i}}(t)}{\partial\mathbf{y}_{_{M,i}}(t)}\boldsymbol{\Phi}_{_{MS,i}}(t,t_{i}).\label{eq:couplingTerm}
\end{align}
Similarly, we require a formulation for $\boldsymbol{S}_{_{MS,i}}$, given by
\textcolor{black}{
	\begin{align}
	&\frac{d \mathbf{S}_{_{M,i}}(t)}{dt}=\frac{\partial\dot{\mathbf{y}}_{_{M,i}}(t)}{\partial\mathbf{y}_{_{S}}(t)}\mathbf{S}_{_{S}}(t)+
	\frac{\partial\dot{\mathbf{y}}_{_{M,i}}(t)}{\partial\mathbf{y}_{_{M,i}}(t)}\mathbf{S}_{_{M,i}}(t)+\frac{\partial\dot{\mathbf{y}}_{_{M,i}}(t)}{\partial\mathbf{p}}.\label{eq:coupledSensitivity}
	\end{align}
}
This completes the required formulation for the differential equations governing the evolution of Eqs. (\ref{eq:totalPhi}) and (\ref{eq:totalS}). From the single-arc propagation, we retrieve $\mathbf{y}_{_{S}}$, $\mathbf{\Phi}_{_{SS}}$ and $\mathbf{S}_{_{S}}$, appearing in Eqs (\ref{eq:couplingTerm}) and (\ref{eq:coupledSensitivity}), and use as a given when solving the multi-arc dynamics and variational equations.

The advantage of this  approach, with two separate integrations to fully populate the coupled $\boldsymbol{\Phi}$ and $\mathbf{S}$ matrices, is that different numerical settings may be used for the single- and multi-arc segment. In particular, for the case of coupled natural body and spacecraft dynamics estimation, one will typically require a much smaller time-step for propagating the spacecraft than for the natural bodies (as well as possibly a different integrator).

\subsubsection{ Formulation for the JUICE mission} \label{sec:coupledMethodJuice}
In Section \ref{sec:coupledVariationalEquations} we presented our general framework for propagating coupled single- and multi-arc variational equations. We now discuss specific details of the formulation for the JUICE mission. Our single-arc state vector is defined  in a planetocentric reference frame as
\begin{align}
\mathbf{y}_{_{S}}(t)=\begin{pmatrix}\mathbf{x}_{1}^{(0)}(t)\\\mathbf{x}_{2}^{(0)}(t)\\\mathbf{x}_{3}^{(0)}(t)\\\mathbf{x}_{4}^{(0)}(t)\end{pmatrix},
\end{align}
where the index 0 refers to Jupiter, and indices 1,2,3,4 correspond to Io, Europa, Ganymede and Callisto, respectively, following \cite{dirkx2016}.

Only the spacecraft's dynamics are solved for in arc-wise manner, such that the multi-arc state vector for arc $i$ can simply be written as 
\begin{align}
\mathbf{y}_{_{M,i}}(t)=\mathbf{x}_{\mathrm{sc},i}^{(j_{i})}(t).
\end{align}
We use `$\mathrm{sc}$' to denote properties relating to the JUICE spacecraft, while $j_{i}$ designates the index $j$ of the central body during arc $i$. The reference frame origin is selected as the moon where a flyby is performed during the flyby phase (Europa, Ganymede or Callisto), and as Ganymede during the orbital phase. 

Solving the coupled variational equations provides solutions for the derivatives $\frac{\partial\mathbf{x}_{\mathrm{sc},i}^{(j_{i})}}{\partial *}$, which describe changes in the moon-centered state of the JUICE spacecraft. However, to evaluate our design matrix $\mathbf{H}$ (see Eq. \ref{eq:partialsMatrix}), we need to account for the variations in the \textit{observed} position of the spacecraft, often expressed in an inertial frame (\emph{e.g.} Solar System Barycentre). As a result, the dynamics of the moons influence the observed position of the spacecraft in two distinct manners:
\begin{itemize}
	\item the dynamical contribution, through the bottom-left block of Eq. (\ref{eq:totalPhi}),
	\item the kinematic or \textit{indirect} contribution, through the variations in the moons' states with respect to the reference frame used for the observed spacecraft's motion. 
\end{itemize}

This methodology automatically allows the incorporation of parameters that directly influence both the spacecraft's and moon's dynamics. Principally, this concerns the moons' spherical harmonic coefficients. Consistently propagating $\mathbf{S}(t)$ for the full system ensures that the covariance of the moons' initial states is robustly propagated to later epochs (see Section \ref{sec:covariancePropagation}). 

\subsection{Decoupled single- and multi-arc estimation} \label{sec:decoupledModel}

To complement the description of the coupled estimation method in Section \ref{sec:coupledModel}, the decoupled strategy is now discussed. As this approach does not differ from textbook formulations \textcolor{black}{\citep[\textit{e.g.}][]{montenbruckGill2000, milaniGronchi2010}}, less details are provided and we directly address the JUICE case specifically. For our comparative analysis, it is however crucial to make both the decoupled and coupled formulations explicit, to highlight their main differences. 

\subsubsection{General principle} \label{sec:generalPrincipleDecoupled}

The decoupled estimation is performed in two separate steps, as shown in Figure \ref{fig:decoupledModel}. The spacecraft's and natural bodies' dynamics are first solved for concurrently, as in the coupled case, but in a multi-arc manner. Only the dynamical coupling between the spacecraft and the central body is thus accounted for in this estimation step (while all dynamical couplings are included to propagate the moons' states, see Section \ref{sec:moonDynamics}). Since the natural bodies' states are independently estimated for each arc, the adopted dynamical model need not be consistent over long timescales. 

This first estimation step therefore provides arc-wise estimated states for the central bodies. These so- called \textit{normal points} are then used as observables in a second step, which aims at reconstructing the natural bodies' dynamics on a more global scale. More precisely, a normal point is defined as the central moon's cartesian state components \textcolor{black}{(vector of size 6)} and associated covariances \textcolor{black}(6-by-6 matrices), determined with respect to Jupiter at the time of closest approach. The covariances $\mathbf{P_{qq}}$ for the arc-wise initial states, resulting from the first estimation step, determine the weights $\mathbf{W}$ (see Eq. \ref{eq:covariance}) assigned to each normal point in the second step. \textcolor{black}{The matrix $\mathbf{W}$ is thus not exactly diagonal in this particular case (see Section \ref{sec:covarianceAnalysis}). It instead shows non-zeros, diagonally-centered, 6-by-6 blocks containing the 6-by-6 normal points' covariances.} The entire two-step decoupled estimation process is depicted in Figure \ref{fig:decoupledModel}, using JUICE flybys as an example.

While the coupled model estimates all parameters concurrently, different sets of estimated parameters are defined for the two steps of the decoupled method. An obvious example are the spacecraft's states, which are determined in a multi-arc manner in the first step but are absent from the second step, when reconstructing the global solution for the moons (see Figure \ref{fig:decoupledModel}). \textcolor{black}{As previously mentioned, it must be noted that the first step of the decoupled model only estimates the state of the central moon $j$ when determining the normal point for a flyby around that moon. The state uncertainties of the other moons are not accounted for, and our decoupled estimation strategy might thus yield slightly too optimistic formal errors for the arc-wise state of moon $j$. However, as a verification, we ran an additional analysis for the JUICE test case including the other moons' states as consider parameters in the normal points determination process. As an indication, we provide the results obtained for two of the JUICE flybys in \ref{appendix:nonCentralMoonsConsiderParameters}. This verification showed that these uncertainties have a negligible impact on the normal points solution when using tracking arcs of 8 hours only around each flyby (see Section \ref{sec:estimationSettings}). This assumption should however be revisited if longer tracking arcs were to be considered for the JUICE flybys, as the influence of the state uncertainties for the non-central moons is then expected to increase.} A more complete discussion on the estimated parameters for our JUICE analysis is provided in Section \ref{sec:estimationSettings} (see Table \ref{tab:parameters}).

\subsubsection{Decoupled variational equations for the JUICE mission}

As described in Section \ref{sec:generalPrincipleDecoupled}, the spacecraft's and moon's dynamics are first reconstructed in a multi-arc fashion. For each arc $i$, the initial translational state to be estimated is thus defined as
\begingroup
\renewcommand*{\arraystretch}{1.7}
\begin{align}
\mathbf{y}_{_{M,i}}(t_i) = \begin{pmatrix}
\mathbf{x}^{(0)}_{j_i,i}(t_i) \\ 
\mathbf{x}_{\mathrm{sc},i}^{(j_{i})}(t_i)
\end{pmatrix} \label{eq:decoupledMultiArcState}
\end{align}
\endgroup
where $j_i$ again refers to the index of the central moon for arc $i$.

In practice, all arcs sharing the same central moon are combined, to allow some dynamical parameters to be estimated globally alongside the arc-wise states (\textit{e.g.} gravity field coefficients of the central moon). For each moon $j$, the full initial state is thus built by concatenating the corresponding multi-arc states, as follows:
\begingroup
\renewcommand*{\arraystretch}{1.7}
\begin{align}
\mathbf{y}_j = \begin{pmatrix}
\mathbf{x}^{(0)}_{j}(t_1) \\
\mathbf{x}^{(j)}_\mathrm{sc,1}(t_1)\\
\mathbf{x}^{(0)}_{j}(t_2) \\
...
\\
\mathbf{x}^{(j)}_{sc,N_j}(t_{N_j})
\end{pmatrix}, \label{eq:decoupledInitialState}
\end{align}
\endgroup
with $N_j$ the number of arcs with moon $j$ as central body.

The arc-wise state transition matrix $\mathbf{\Phi}_{i}(t,t_i)$ can be derived from Eq. (\ref{eq:decoupledMultiArcState}) as 
\begin{align}
\mathbf{\Phi}_i(t,t_i) &= \frac{\partial \mathbf{y}_{_{M,i}}(t)}{\partial\mathbf{y}_{_{M,i}}(t_i)}, t\in[t_{i},\tilde{t}_{i}] \\
&= \begin{pmatrix}
\mathbf{\Phi}_{j_i}(t;t_i) & \mathbf{0}_{6,6}\\
\mathbf{\Phi}_{\mathrm{sc},j_i}(t,t_i) & \mathbf{\Phi}_{\mathrm{sc}}(t,t_i)
\end{pmatrix}. \label{eq:arcWisePhiDecoupled}
\end{align}
Eq. (\ref{eq:arcWisePhiDecoupled}) shows some similarities with Eq. (\ref{eq:totalPhi}), but also clearly highlights major differences between the coupled and decoupled formulations. In particular, $\mathbf{\Phi}_{\mathrm{sc},j_i}(t,t_i)$ also represents a coupling term, but expressed in a multi-arc fashion and with respect to the central moon $j_i$ only:
\begin{align}
\mathbf{\Phi}_{\mathrm{sc},j_i}(t;t_i) = \frac{\partial \mathbf{y}_{\mathrm{sc}}(t)}{\partial \mathbf{y}_{j_i}(t_i)}, 
\end{align}
as opposed to Eq. (\ref{eq:coupledSTM}).

The variational equations provided above apply to the first, arc-wise estimation step of the decoupled strategy (see Section \ref{sec:generalPrincipleDecoupled}). The second phase, in which a single-arc estimation is performed to reconstruct the long-term dynamics of the Galilean moons, follows the regular single-arc approach. The associated variational equations are therefore not detailed in this paper.

\subsubsection{\textit{A priori} knowledge strategy} \label{sec:aPrioriKnowledge}

As shown in Equation \ref{eq:covariance}, prior knowledge is accounted for in the estimation by means of the \textit{a priori} covariance matrix $\mathbf{P}_{\mathbf{q}\mathbf{q},0}$. Appropriate \textit{a priori} values for all estimated parameters, referred to as \textit{default} \textit{a priori} covariances, are further discussed in Section \ref{sec:estimationSettings} and are combined in a diagonal matrix $\mathbf{P}_0$ (shortened notation for $\mathbf{P}_{\mathbf{q}\mathbf{q},0}$). 

For the decoupled case in particular, the moons' arc-wise state solutions first determined at the beginning of each flyby strongly depend on these \textit{a priori} constraints (see results in Section \ref{sec:sensitivityAPriori}). Using the same \textit{default a priori} values for all arc-wise moon states, derived from the existing ephemerides solutions, would be a rather conservative approach. It indeed neglects the iterative improvement achievable by progressively including more flybys in the estimation. \textcolor{black}{Even if the observations processed by the estimation remain the same, some additional information is incorporated in the multi-arc model to improve the solution, namely that the arc-wise state solutions for a given moon $j$ are not completely independent from one another. They indeed belong to a single body's trajectory and should thus be dynamically consistent. Such an update strategy for the \textit{a priori} contraints on the moons' states can be compared to the multi-arc constrained approach for the spacecraft's orbit determination \citep[\textit{e.g.}][]{alessi2012}, but applied to the moons' arc-wise states instead of the spacecraft's.} \textcolor{black}{It must be noted that using this \textit{a priori} update strategy introduces some correlations between the arc-wise state components of moon $j$ (\textit{i.e.} between the different normal points determined for this moon). The off-diagonal blocks of the weight matrix $\mathbf{W}$ are therefore not filled with zeros anymore.}

Focusing on the $N_j$ arcs with moon $j$ as central body, more realistic \textit{a priori} covariances can be derived for arc $k$ by propagating the covariance obtained for arc $k-1$ up to the beginning of arc $k$. This propagated covariance is denoted as $\mathbf{P}_0^{k \rightarrow k+1}$ in the following. Some state components may nonetheless be poorly constrained by the previous arc's estimation, thus yielding unrealistically large \textit{a priori} errors in certain directions. The \textit{a priori} matrix $\mathbf{P}_0^k$ for arc $k$ is thus built as a combination of the default and propagated \textit{a priori} covariance matrices, as follows:
\textcolor{black}{\begin{align}
	\left(\mathbf{P}_0^{k}\right)^{-1} &=  \left(\mathbf{P}_0\right)^{-1} + \left(\mathbf{P}_0^{k-1 \rightarrow k}\right)^{-1} \label{eq:updatedAPriori} \\
	&=  \left(\mathbf{P}_0\right)^{-1} + \left(\mathbf{\Phi}_j(t_{k-1},t_{k})\mathbf{P}_0^{k-1}\mathbf{\Phi}_j(t_{k-1},t_{k})^{T}\right)^{-1}, \nonumber 
	\end{align}
}
where $\mathbf{\Phi}_j(t_{k-1},t_{k})$ is the state transition matrix for moon $j$, computed from the start of arc $k-1$ to the beginning of the current arc $k$. This propagation scheme is initialised with the \textit{default} \textit{a priori} matrix \textcolor{black}{(so $\mathbf{P}_0^{0\rightarrow1} = \mathbf{0}$, \textit{i.e.} matrix filled with zeros)}.

\begin{figure}[tbp]
	\centering
	\includegraphics[width=0.5\textwidth]{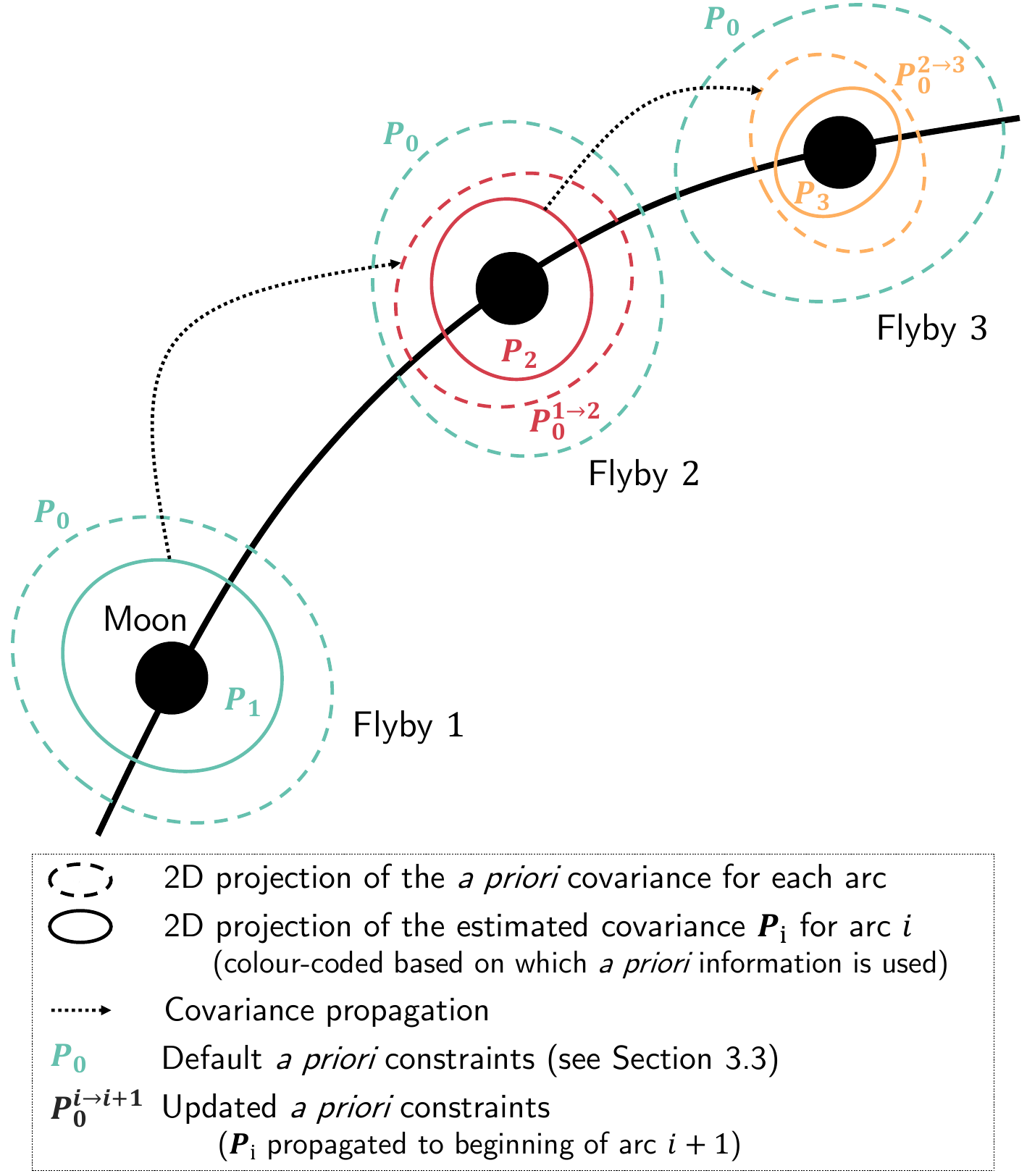}
	\caption{Schematic representation of the iterative strategy for the \textit{a priori} covariances, applied to the first step of the decoupled method (\textit{i.e.} normal points determination). The dashed ellipses represent the two-dimensional projection of the \textit{a priori} covariance for the arc-wise position of the moon. The solid ellipses display the two-dimensional projection of the estimated covariance after inversion, for each arc. These solid covariance ellipses are colour-coded to indicate which \textit{a priori} values are used. The \textit{default a priori} information $\mathbf{P}_0$ is represented in green. A dashed black arrow indicates a covariance propagation (\textit{i.e.} update of the \textit{a priori} covariance in this case). This representation only illustrates the update principle and is not to scale.}
	\label{fig:aPrioriDecoupledModel}
\end{figure}

Iterating on the \textit{a priori} knowledge for the moons' arc-wise states requires to run the first step of the decoupled estimation multiple times, gradually increasing the number of arcs being processed. The final outcome of the multi-arc estimation (\textit{i.e.} normal points and global parameters' estimates, see Section \ref{sec:decoupledModel}) is reached when all $N_j$ arcs associated with moon $j$ are included. This process is schematically summarised in Figure \ref{fig:aPrioriDecoupledModel}. 

It must be stressed that, in the strategy described above, we only propagate the covariances between moon $j$'s own state components from arc $k-1$ to the beginning of arc $k$. We thus neglect the influence that uncertainties in other moons' states could have on the propagated \textit{a priori} covariance for moon $j$. Discarding the contribution of the other moons is consistent with the philosophy of the decoupled estimation, in which only the central moon's state is determined for each arc. It should however be noted that the \textit{a priori} used for the normal points might therefore be slightly too optimistic.

The impact of the \textit{a priori} information on the parameters solution, and especially the effect of the above updating strategy for the arc-wise states, will be further investigated and discussed in Section \ref{sec:sensitivityAPriori}. For each estimated parameter, the contribution of the \textit{a priori} information to the solution $c_\mathbf{q}$ can be evaluated as follows \citep[\textit{e.g.}][]{floberghagen2001}:
\begin{align}
c_\mathbf{q} = \mathbf{I} - \mathbf{P}\text{ }\mathbf{P}^{-1}_{0}, \label{eq:contributionObservations}
\end{align}
where $\mathbf{I}$ is to the identify matrix, while $\mathbf{P}$ and $\mathbf{P}^{-1}_{0}$ refer to the final and \textit{a priori} covariance matrices, respectively ($\mathbf{P}_{\mathbf{q}\mathbf{q}}$ and $\mathbf{P}^{-1}_{\mathbf{q}\mathbf{q},0}$ in Eq. \ref{eq:covariance}). A $c_\mathbf{q}$ equal to 1 indicates that the parameter's estimation relies entirely on the observations, while a value of 0 means that it is based on \textit{a priori} information.

\subsection{Scope of the comparative analysis} \label{sec:scopeComparativeAnalysis}

As mentioned in Section \ref{sec:introduction}, we limit ourselves to a covariance analysis in this study. We compare the performances of the decoupled and coupled models by analysing the formal errors and correlations obtained in both cases. It is important to stress that the coupled model, by concurrently accounting for all dynamical couplings and sensitivities, directly maps the simulated observations set to the estimated parameters' covariances. Assuming that  the fidelity of both our dynamical and measurement error models is sufficient, the resulting formal errors and correlations are therefore considered to provide a good statistical representation of the estimation solution, while this is not directly true for the decoupled method. Our study characterises how much the solution obtained by decoupling the spacecraft's and moons' state estimation departs from the covariances given by the coupled approach, regarded as statistically consistent. 

For any estimation, true errors obtained from real data after completing the iterative least-squares estimation are however larger than the formal errors provided by a covariance analysis. Differences between true and formal errors originate from non-white measurement noise, as well as inaccuracies in the models used for the spacecraft's and planetary system's dynamics. For the JUICE mission data analysis, this situation may even be more severe than for any previous natural satellite's ephemeris determination, due to the much better data quality and subsequent higher requirements on dynamical modelling. 

In practice, both the decoupled and coupled methods are limited by the dynamical model fidelity, so that the true errors would be larger than formal ones in the two cases. The goal of this study is to determine at which point and to what extent the coupled estimation would be beneficial for ephemerides determination, as well as to quantify the dynamical model requirements to achieve this (see Section \ref{sec:introduction}). Dynamical mismodelling would nonetheless influence the decoupled and coupled solutions differently, and represent a major challenge for the applicability of the coupled model in particular. These modelling issues will therefore be discussed in more detail in Section \ref{sec:discussion}.

\section{Dynamical and Observation Models}
\label{sec:dynamicsAndObservations}

In the following section, we discuss the settings and models used for our simulated covariance analysis for Galilean satellites' ephemerides from JUICE tracking data. The spacecraft's and moons' dynamical models are summarised in Sections \ref{sec:spacecraftDynamics} and \ref{sec:moonDynamics}, respectively, while the estimation and observations settings are described in Section \ref{sec:estimationSettings}. 

\subsection{Spacecraft Dynamics}
\label{sec:spacecraftDynamics}

When propagating the dynamics of the spacecraft during arc $i$, the following accelerations were taken into account:
\begin{itemize}
	\item spherical harmonic acceleration of the central moon $j_{i}$, expanded up to degree $l_{j}$ and order $m_{j}$ (with $l_{1}/m_{1}$ = 2/2, $l_{2}/m_{2}$ = 4/4, $l_{3}/m_{3}$ = 12/12, $l_{4}/m_{4}$ = 6/6). Higher degrees might be accessible from JUICE data, especially for Ganymede, but were purposely not included in our analysis, which primarily focuses on ephemerides determination.
	\item point-mass acceleration of moon $k$, for each $k\in \{1,2,3,4\}$, with $k\neq j_{i}$ 
	\item spherical harmonic acceleration of Jupiter, expanded up to degree $l_{0} = 8$ and order 0 (zonal terms only),
	\item point-mass accelerations exerted by Saturn and the Sun,
	\item cannonball radiation pressure acceleration due to the Sun's radiation,
	\item arc-wise constant (in RTN frame) empirical acceleration, representing errors in the accelerometer calibration.
\end{itemize}

We adopted the same models for the environment (gravity fields, ephemerides, \emph{etc.}) as \cite{dirkx2016}, with a number of exceptions: we used the Jupiter gravity field from \cite{iess2018}, and CReMA 5.0 for the JUICE orbit\footnote{https://www.cosmos.esa.int/web/spice/spice-for-juice}, \textcolor{black}{released} by ESA in the form of Spice kernels \citep{acton1996}. 

\subsection{Moon Dynamics}
\label{sec:moonDynamics}
When propagating the dynamics of the Galilean moons, we used similar models as \cite{lainey2004,dirkx2016}, taking into account
\begin{itemize}
	\item the mutual spherical harmonic acceleration between Jupiter and each moon $j$, with the gravity field of Jupiter expanded up to degree 8 and order 0, and that of the moons to degree and order 2,
	\item the mutual spherical harmonic accelerations between all moons $j$ and $k$, with the fields of the bodies expanded up to degree and order 2, 
	\item the point-mass accelerations due to Saturn and the Sun,
	\item the acceleration exerted on each moon $j$ due to tidal dissipation in Jupiter forced by moon $j$,
	\item the acceleration on each moon $j$ due to tidal dissipation in moon $j$ forced by Jupiter. The influence of the tides raised by body $k$ on body $j$ on the system's dynamics was modelled by time-varying corrections applied to the spherical harmonics expansion of the body $j$' gravity field, as done in \citep[\textit{e.g.}][]{dirkx2016,dirkx2017}.
\end{itemize}

\subsection{Estimation Settings}
\label{sec:estimationSettings}

As the tracking configurations differ between the flyby and orbital phases of the JUICE mission, different estimation settings were used. An 8-hour tracking arc was defined for each flyby, centered at the time of closest approach. For the orbital phase, we simulated 8 hours of tracking per day. In practice, the JUICE spacecraft will be tracked from three stations of the European Space Tracking network (ESTRACK), the main one being Malargue, which is as of yet the only one enabling both X- and Ka-band tracking. However, we assumed that the other two will also be able to handle Ka-band tracking by the time the JUICE spacecraft arrives in the Jovian system. We thus considered 8 hours per day of almost continuous tracking, except during occultations or for elevations lower than 15 deg \citep[as in \textit{e.g.}][]{diBenedetto2021, magnanini2021}. 

Tracking arcs of two days, separated by three days without tracking, were used as the nominal tracking configuration for the orbital phase. Nonetheless, the sensitivity of the estimation solution to these tracking settings was investigated by considering one day- and one week-arcs (results are presented in Section \ref{sec:sensitivityTrackingSchedule}). The three days interval between two tracking arcs was merely used to reduce the computational load of our simulations. \textcolor{black}{We verified that adding some buffer between tracking arcs did not affect the resulting formal uncertainties and, most importantly, the way the coupled solution compares to the decoupled one.} 

For each arc, we simulated both Doppler and range observables \textcolor{black}{which are measurements, in the line of sight direction, of the spacecraft's position and velocity with respect to a ground station, respectively}. Doppler observables were modelled with a noise level of 15 $\mu$m/s at an integration time of 60 s, while range observables have a noise level of 20 cm. \textcolor{black}{This is quite precise but should actually be a conservative value, given the 1 cm range accuracy achieved by the BepiColumbo mission \citep[\textit{e.g.}][]{genova2021}.} For selected passes, as was done by \cite{dirkx2017}, we also simulated VLBI observables 
(lateral position of the target spacecraft) following methodology described by \cite{pogrebenko2004} and \cite{duev2016}, with a noise level of 0.5 nrad. Doppler data were generated as unbiased, while we included arc-wise biases for both the range and VLBI observables. It should be noted that range and Doppler data are obtained in a topocentric frame, while VLBI observations are measured in the ICRF. For both the flyby and orbital phase, observations are subject to constraints on ground station visibility (occultation, Sun angle). 

When presenting and discussing our results in the rest of this paper, the estimated states will generally be expressed in the RTN frame: the x-axis points from the central body towards the spacecraft or moon, the z-axis is aligned with the normal to the orbital plane and the y-axis completes the reference frame. In the following, they are referred to as the radial, tangential and normal directions, respectively.

In our simulations, we estimated the following set of parameters:
\begin{itemize}
	\item arc-wise JUICE initial states $\mathbf{x}^{(j_{i})}_{\mathrm{sc}}(t_{i})$ ($i=1...N$), with \textit{a priori} uncertainty of 5 km and 0.5 m/s on
	position and velocity components, respectively.
	\item \textcolor{black}{global or arc-wise} Galilean moons' initial states $\mathbf{x}^{(0)}_{j}(t_{0}/t_{i})$ ($j=1..4)$. The \textit{a priori} uncertainty in position was set to 15 km in the three RTN directions. For the velocity components, we used the differences between the latest IMCCE and JPL ephemerides (NOE-5-2021\footnote{https://ftp.imcce.fr/pub/ephem/satel/NOE/JUPITER/} and JUP365\footnote{https://ssd.jpl.nasa.gov/sats/ephem/}, respectively) as a conservative \textit{a priori}. \textcolor{black}{These \textit{a apriori} values are provided in Table \ref{tab:aPrioriVelocities}.}
	\textcolor{black}{\item gravitational parameters of Galilean moons $\mu_j$ ($j=1..4$), using the \textit{a priori} uncertainties provided in \cite{schubert2004} and reported in Table \ref{tab:aPrioriGravity}.}
	\item gravity field coefficients of Galilean moons $C_{lm}^{(j)}$, $\mathbf{S}_{lm}^{(j)}$, up to degree and order 2, 4, 12 (6 when considering the flyby phase only) and 6, for Io, Europa, Ganymede and Callisto respectively. \textcolor{black}{As \textit{a priori} constraints, we used the formal uncertainties by \cite{schubert2004} for $\bar{C}_{20}$ and $\bar{C}_{22}$, which are given in Table \ref{tab:aPrioriGravity}. We applied Kaula's rule with $K=10^{-5}$ for the remaining gravity field coefficients \citep[$\sigma = K / l^2$,][]{kaula1966}}.
	\item arc-wise accelerometer bias calibration factors $\mathbf{c}_{i}$, with the \emph{a priori} constraint set to $10^{-7}$ m$\cdot$s$^{-2}$ \citep[10 times larger than in][]{cappuccio2020}.
	\item arc-wise biases for range observables, with an \textit{a priori} uncertainty fixed to 0.25 m.
	\item arc-wise biases for VLBI observables. We set the bias constraint at 0.5 nrad in both right ascension and declination \citep{charlot2020}.
\end{itemize}

\begin{table*}[tbp]
	\caption{\textcolor{black}{\textit{A priori} constraints for the velocity components of the Galilean moons' states, expressed in the RTN  reference frame. These \textit{a priori} values are computed as the differences between the NOE-5-2021 and JUP365 ephemerides, averaged over the JUICE mission timeline.}}
	\label{tab:aPrioriVelocities}
	\centering
	\begin{tabular}{l c c c }
		\hline
		&  \textbf{Radial [m/s]} & \textbf{Tangential [m/s]} & \textbf{Normal [m/s]}  \\ \hline
		Io & 0.98  & 0.14   & 0.72 \\
		Europa & 0.35  & 0.10  & 0.74 \\
		Ganymede & 0.21  & 0.08 &  0.32  \\
		Callisto & 0.16  & 0.07 &  0.10  \\
		\hline
	\end{tabular}
\end{table*}

\begin{table}[tbp]
	\caption{\textcolor{black}{\textit{A priori} constraints for gravitational parameters, normalised $\bar{C}_{20}$ and $\bar{C}_{22}$ coefficients for the four Galilean moons. The values are retrieved from \cite{schubert2004}.}}
	\label{tab:aPrioriGravity}
	\centering
	\begin{tabular}{l c c c }
		\hline
		&  $\mathbf{\mu}$ [$\mathbf{{km/s^2}}$] & $\mathbf{\bar{C}_{20}}$ \textbf{[-]} & $\mathbf{\bar{C}_{22}}$ \textbf{[-]} \\ \hline
		Io & 0.02 &  $\mathrm{2.7\cdot 10^{-6}}$ & $\mathrm{0.8\cdot 10^{-6}}$ \\
		Europa & 0.02 & $\mathrm{8.2\cdot 10^{-6}}$  & $\mathrm{2.5\cdot 10^{-6}}$ \\
		Ganymede & 0.03 & $\mathrm{2.9\cdot 10^{-6}}$ &  $\mathrm{0.87\cdot 10^{-6}}$  \\
		Callisto & 0.01 & $\mathrm{0.8\cdot 10^{-6}}$ &  $\mathrm{0.3\cdot 10^{-6}}$  \\
		\hline
	\end{tabular}
\end{table}

Table \ref{tab:parameters} specifies whether a parameter is to be estimated globally or in an arc-wise manner. It highlights important differences between the two estimation methods, but also between the two steps of the decoupled approach. It must be stressed that the moons' gravity field coefficients are only included in the second step of the decoupled approach to account for the influence of uncertainties in the moons' gravity fields on the propagated state solutions (see Eq. \ref{eq:propagatedCovariance}). This is merely a way to avoid obtaining too optimistic formal errors because part of the uncertainties sources would be omitted. \textcolor{black}{The \textit{a priori} values for these coefficients are directly taken from the formal errors obtained after the first estimation step, and the gravity field solutions are actually not improved further by the second step, compared to these \textit{a prioris}.} 

\textcolor{black}{As shown by \cite{dirkx2016,dirkx2017}, the influence of Jupiter's state and gravity field uncertainties on the estimation results was considered negligible in the post-Juno era \citep{durante2020}, and these parameters were therefore not determined in our simulations. Tidal dissipation parameters were also excluded from the list of parameters to estimate in this preliminary study, keeping the focus of our analysis primarily on state estimation methods and on the resulting solutions for both the spacecraft and the moons.} 

\begin{table}[tbp]
	\caption{Detailed description of the estimated parameters sets, for both the coupled and decoupled estimation approaches. As mentioned in Section \ref{sec:generalPrincipleDecoupled}, the parameters change between the first and second steps of the decoupled method. NI stands for `not included'.}
	\label{tab:parameters}
	\centering
	\begin{tabular}{l c c c }
		\hline
		&  \textbf{Coupled} & \multicolumn{2}{c}{\textbf{Decoupled}} \\ &   &  1st step & 2nd step \\ \hline
		
		JUICE's states & \textcolor{black}{arc-wise} & \textcolor{black}{arc-wise} & NI   \\
		Moons' states & \textcolor{black}{global} & \textcolor{black}{arc-wise}  & \textcolor{black}{global} \\
		Moons' gravity coef. & \textcolor{black}{global} & \textcolor{black}{global} & \textcolor{black}{global} \\
		Accelerometer biases & \textcolor{black}{arc-wise} & \textcolor{black}{arc-wise} & NI \\
		
		Range biases& \textcolor{black}{arc-wise} & \textcolor{black}{arc-wise} & NI \\
		VLBI biases & \textcolor{black}{arc-wise} & \textcolor{black}{arc-wise} & NI \\
		\hline
	\end{tabular}
\end{table}

\section{Results} \label{sec:results}

This section presents the results of our comparative covariance analyses, performed with both the coupled and decoupled estimation models (Sections \ref{sec:coupledModel} and \ref{sec:decoupledModel}). We first only considered the flyby phase, before including the orbital phase. The results obtained in the two configurations are presented in Sections \ref{sec:resultsFlybyPhase} and \ref{sec:resultsOrbitalPhase}, respectively.

It should first be highlighted that the estimation problem is very close to being ill-posed, with extremely high condition number for the normal equations (Eq. \ref{eq:covariance}). The exact values of the formal errors provided in the coming section should thus be treated cautiously. \textcolor{black}{However, it must be noted that, as a verification, we also performed a deterministic least-squares estimation for the coupled case, to bring confidence in the formal uncertainties level (see \ref{appendix:verification}). It proves that our implementation of the lesser documented coupled model is correct, and that the obtained formal errors would be representative of the true errors under the assumptions of a covariance analysis (perfect dynamical and observational models, to be further discussed in Section \ref{sec:discussion}).} Our results therefore remain insightful, especially since we focus on comparing two estimation strategies (and not on absolute error values). 

Given the near ill-posedness of the Galilean moons' state estimation problem, it is worth stressing that the condition number is higher when using the decoupled method, which is an important disadvantage of this approach. When reconstructing the moons' long-term dynamics from the normal points, extremely high correlations between the position and velocity components in the radial and tangential directions even made the estimation problem non-invertible at first. Eventually, only the normal points' positions were therefore added as observables in the second step of the decoupled method (Section \ref{sec:generalPrincipleDecoupled}), to partially eliminate these correlations. \textcolor{black}{In most other analyses, the normal points also include the central moon's position only \citep[\textit{e.g.}][]{durante2019,diruscio2020,diruscio2021}}

\subsection{Flybys phase only} \label{sec:resultsFlybyPhase}

This section presents the covariance analysis results obtained from observations simulated over the JUICE flyby phase only. We first discuss and compare the resulting formal errors in Galilean moons' states (Section \ref{sec:stateErrorsFlybyPhase}), and in gravity field coefficients (Section \ref{sec:gravityErrorsFlybyPhase}). The sensitivity of the estimation solutions to the \textit{a priori} covariances for the moons' initial states is then investigated in Section \ref{sec:sensitivityAPriori}.   

\subsubsection{State estimation} \label{sec:stateErrorsFlybyPhase}

\begin{figure*}
	\centering
	\makebox[\textwidth][c]{\includegraphics[width=1.0\textwidth]{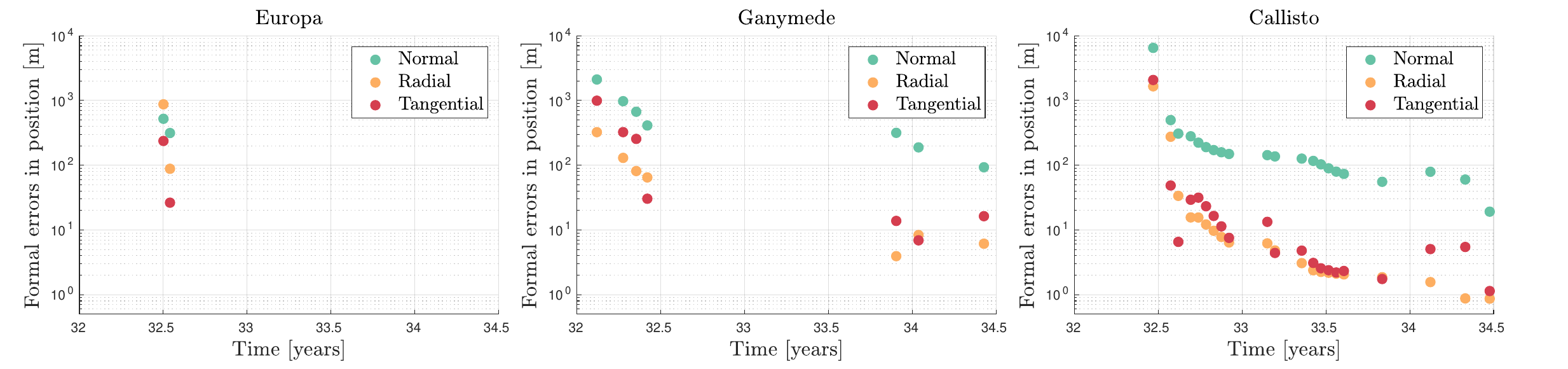}}
	\caption{Formal errors in position obtained \textcolor{black}{after the first step of the decoupled estimation} for each normal point (\textit{i.e.} each flyby represented in Figure \ref{fig:flybysAltitudes}), for the three Galilean moons targeted by the JUICE spacecraft. They correspond to the $1\sigma$ uncertainties as provided by the covariance analysis, and are here expressed in the RTN frame.}
	\label{fig:normalPoints}
\end{figure*}

\begin{figure*}
	\centering
	\makebox[\textwidth][c]{\includegraphics[width=1.0\textwidth]{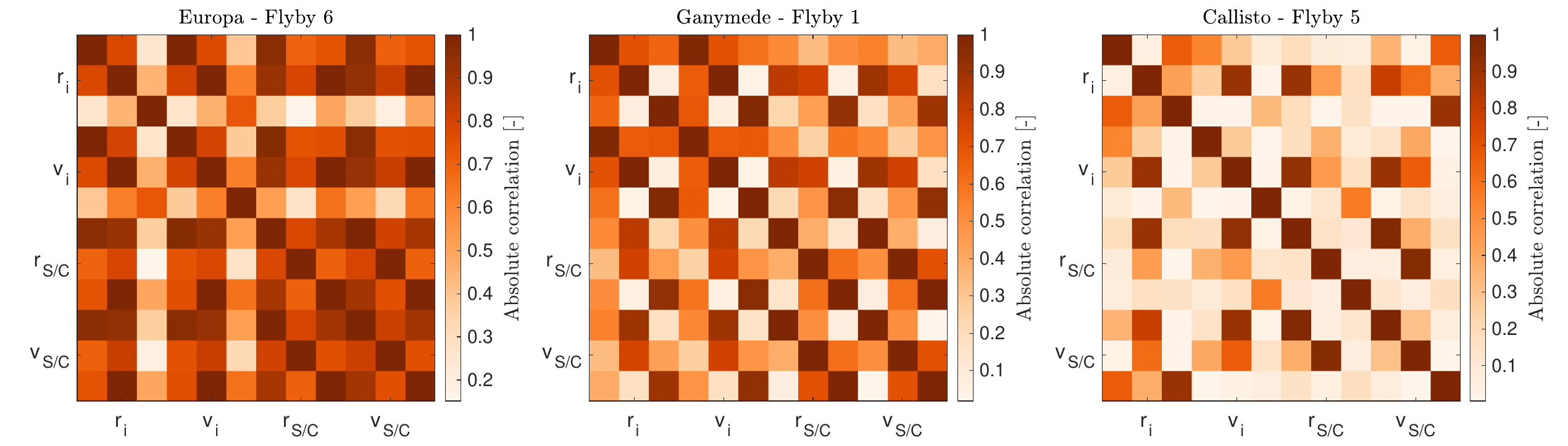}}
	\caption{Absolute correlations obtained when generating the first normal point of each moon (so for each first flyby at Europa, Ganymede and Callisto, respectively). The correlations are computed between the central moon's and spacecraft's state components, both expressed in the RTN frame.}
	\label{fig:normalPointsCorrelations}
\end{figure*}

\textcolor{black}{Focusing on the first step of the decoupled estimation strategy, the uncertainties of the normal points generated for each flyby are displayed in Figure \ref{fig:normalPoints}.} For each moon, the reduction in the normal points' uncertainties as the number of flybys increases is clear.  This is a direct consequence of updating the \textit{a priori} knowledge for the moons' states, ensuring that each arc benefits from the previous ones (Section \ref{sec:aPrioriKnowledge}). Figure \ref{fig:normalPoints} shows that the position of the central moon is much better determined in the radial and tangential directions than it is in the normal direction (\textit{i.e.} out-of-plane). Interestingly, the errors in radial and tangential positions are of similar orders of magnitude, especially for Callisto. This is due to high correlations between the tangential and radial state components. As an indication, Figure \ref{fig:normalPointsCorrelations} shows the absolute correlations obtained when generating the normal points for the first flybys at Europa, Ganymede and Callisto. \textcolor{black}{Callisto's first normal point is much less correlated than Europa's and Ganymede's. Looking at Figure \ref{fig:normalPoints}, it appears that this does not indicate a good normal point determination, but, on the contrary, is due to the fact that the first, relatively high altitude flyby performed at Callisto (see Figure \ref{fig:flybysAltitudes}) does not allow the estimation to significantly improve the normal point determination compared to \textit{a priori} values. The estimated state components for this normal point remain rather uncorrelated (since the \textit{default a priori} covariances assume no correlation between parameters), but the associated formal uncertainties are quite large. The other flybys performed around Callisto, supported by the adopted \textit{a priori} update strategy (see Section \ref{sec:aPrioriKnowledge}), progressively improve the quality of the normal points determination (see Figure \ref{fig:normalPoints}), but also yield much higher normal points' correlations, which then become comparable to ones displayed for Europa and Ganymede in Figure \ref{fig:normalPointsCorrelations}.} 

\begin{figure*} [ht!]
	\centering
	\begin{minipage}[l]{0.49\textwidth}
		\centering
		\makebox[\textwidth][c]{\includegraphics[width=1.0\textwidth]{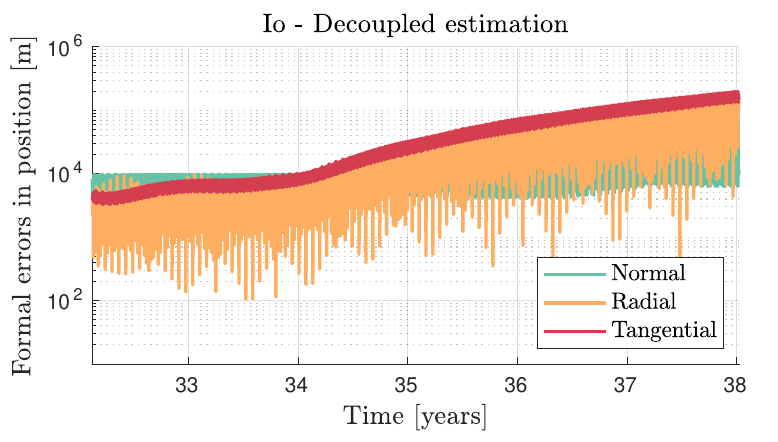}}
		\label{fig:IoDecoupledModelFlybys}
	\end{minipage}
	\begin{minipage}[l]{0.49\textwidth}
		\centering
		\makebox[\textwidth][c]{\includegraphics[width=1.0\textwidth]{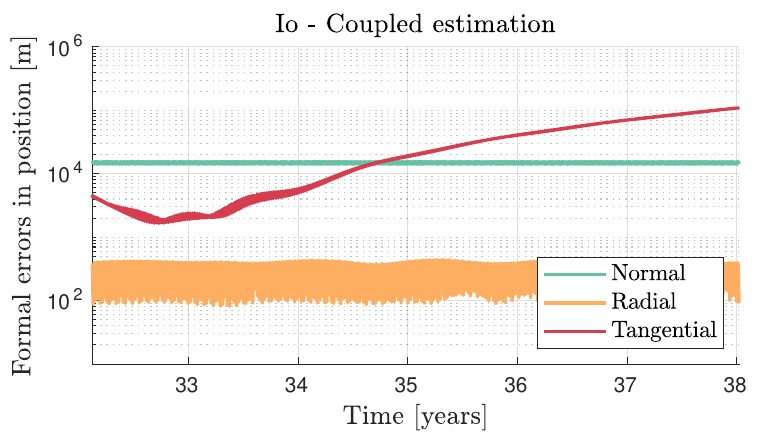}}
		\label{fig:IoCoupledModelFlybys}
	\end{minipage}
	\begin{minipage}[l]{0.49\textwidth}
		\centering
		\makebox[\textwidth][c]{\includegraphics[width=1.0\textwidth]{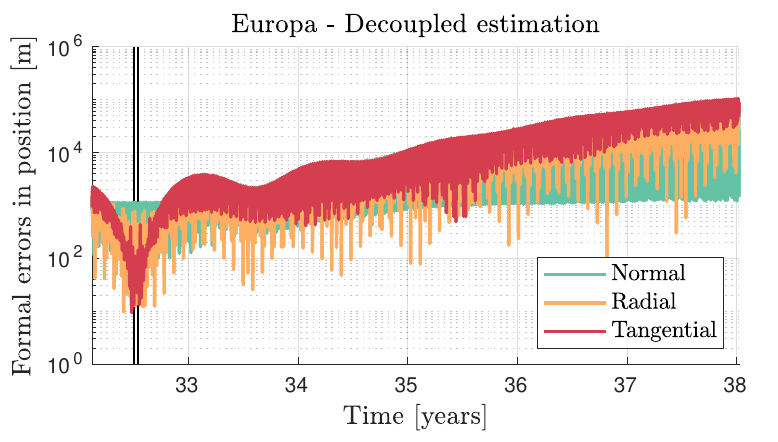}}
		\label{fig:EuropaDecoupledModelFlybys}
	\end{minipage}
	\begin{minipage}[l]{0.49\textwidth}
		\centering
		\makebox[\textwidth][c]{\includegraphics[width=1.0\textwidth]{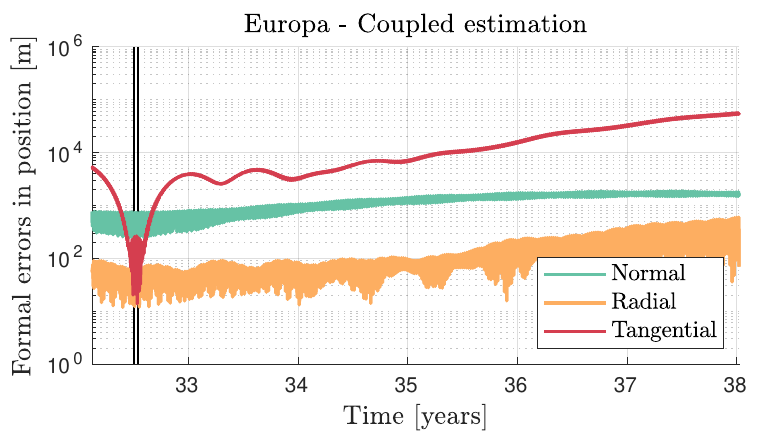}}
		\label{fig:EuropaCoupledModelFlybys}
	\end{minipage}
	\begin{minipage}[l]{0.49\textwidth}
		\centering
		\makebox[\textwidth][c]{\includegraphics[width=1.0\textwidth]{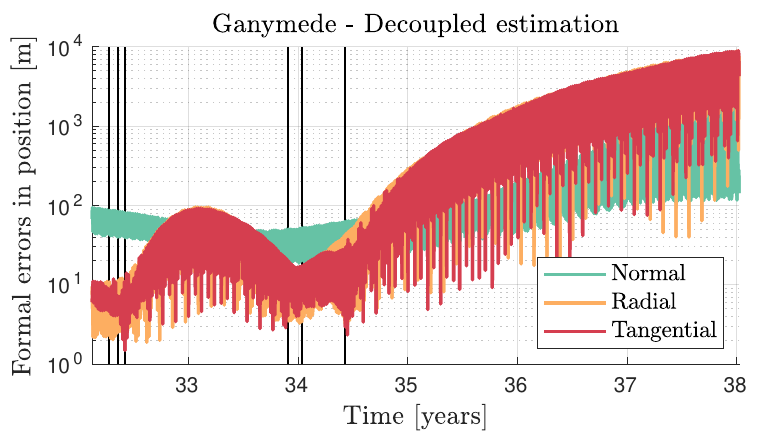}}
		\label{fig:GanymedeDecoupledModelFlybys}
	\end{minipage}
	\begin{minipage}[l]{0.49\textwidth}
		\centering
		\makebox[\textwidth][c]{\includegraphics[width=1.0\textwidth]{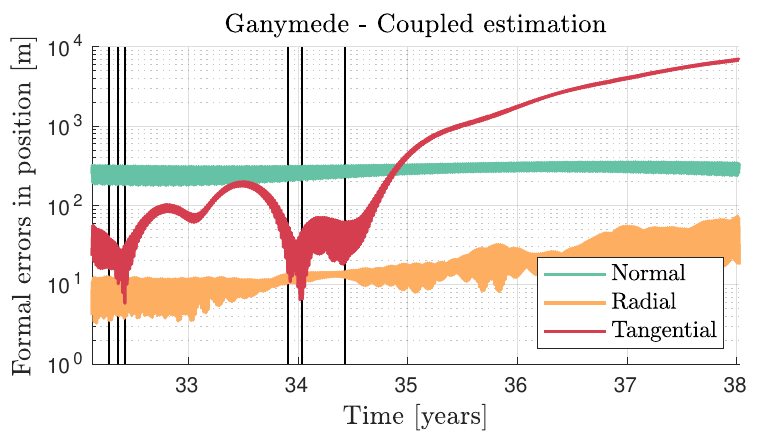}}
		\label{fig:GanymedeCoupledModelFlybys}
	\end{minipage} 
	\begin{minipage}[l]{0.49\textwidth}
		\centering
		\makebox[\textwidth][c]{\includegraphics[width=1.0\textwidth]{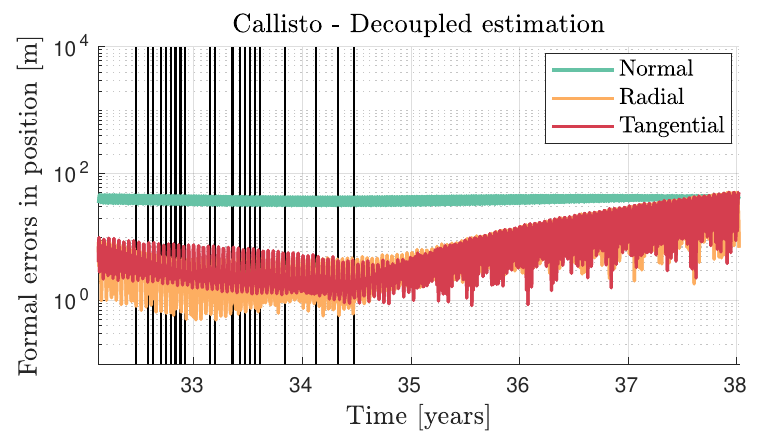}}
		\label{fig:CallistoDecoupledModelFlybys}
	\end{minipage} 
	\begin{minipage}[l]{0.49\textwidth}
		\centering
		\makebox[\textwidth][c]{\includegraphics[width=1.0\textwidth]{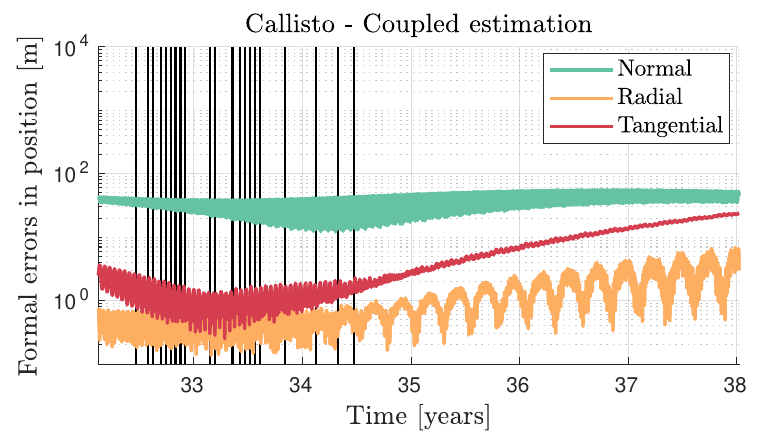}}
		\label{fig:CallistoCoupledModelFlybys}
	\end{minipage}
	\caption{Propagated formal errors in position for the four Galilean moons, obtained with both the decoupled and coupled state estimation methods (left and right sides, respectively). These estimated solutions are based on radiometric data simulated for the JUICE flyby phase only. The black vertical lines indicate when the JUICE flybys occur, for each moon (see Figure \ref{fig:flybysAltitudes}).}
	\label{fig:stateUncertaintiesFlybys}
\end{figure*}

The global solutions for the Galilean moons' dynamics, either reconstructed from the normal points shown in Figure \ref{fig:normalPoints} in the decoupled case, or directly outputted by the coupled model, are displayed in Figure \ref{fig:stateUncertaintiesFlybys}. The $1\sigma$ uncertainties in the moons' states, estimated at the beginning of the flyby phase from all flybys' data, were propagated through the 2030-2038 time period following the methodology presented in Section \ref{sec:covariancePropagation}. The local uncertainty reductions in the propagated solutions clearly indicate when the flybys are performed.

Comparing both solutions in Figure \ref{fig:stateUncertaintiesFlybys}, the coupled method leads to lower formal errors in the moons' radial positions (one or two orders of magnitude lower than in the decoupled case). On the other hand, the uncertainties in the moons' tangential positions are comparable between the two estimation strategies, and can even be locally slightly lower with the decoupled approach. These results follow from the similar formal error levels in the normal points' radial and tangential positions (see Figure \ref{fig:normalPoints}), which translates into uncertainties of comparable orders of magnitude in both directions when reconstructing the global ephemerides solution. On the contrary, the coupled approach is able to more efficiently decorrelate the moon's radial and along-track motion. 

Finally, differences between the coupled and decoupled solutions are not so significant in the normal direction and seem more arbitrary: the decoupled method performs slightly better for Io and Ganymede, while the converse is true for Europa and Callisto. The main difference between the two estimation approaches originates from the decoupled strategy only accounting for the dynamical coupling between the moons in the second step, when trying to reconstruct the moons' dynamics using the kinematic information contained in the normal points. The coupled model, on the other hand, includes all dynamical effects at once. As the strong dynamical coupling between the moons mostly manifests itself in the moons' orbital plane, the results obtained in the normal (\textit{i.e.} out-of-plane) direction are less sensitive to the choice of estimation method. 

While tidal dissipation parameters were intentionally excluded from our comparative state estimation analysis, preliminary insights can still be extrapolated from the expected ephemerides quality. Especially, an accurate determination of the moons' along-track positions is crucial to investigate tidal dissipation effects, \textcolor{black}{through the secular change in mean motion that they induce}. \textcolor{black}{While the decoupled estimation led to slightly lower formal errors for the tangential positions of the moons (see Figure \ref{fig:stateUncertaintiesFlybys}), we should keep in mind that the uncertainties obtained with the coupled model are thought to be more statistically representative, as mentioned in Section \ref{sec:scopeComparativeAnalysis}. It should therefore be highlighted that the determination of the moons' states in the tangential direction might be too optimistic in the decoupled case, possibly translating into lower formal errors for the tidal parameters.} The coupled strategy, on the other hand, may prove beneficial to achieve realistic errors when trying to estimate Jupiter's dissipation at the forcing frequencies of the moons. This would be particularly important to further investigate whether Callisto is caught in a \textcolor{black}{tidal resonance lock} \citep{fuller2016, lainey2020}. It is however essential to stress that, with both models, achieving this tangential uncertainty level will be most challenging, and the results may well be limited by dynamical model error, as further discussed in Section \ref{sec:modellingIssues}.  

\subsubsection{Gravity field estimation} \label{sec:gravityErrorsFlybyPhase}

\begin{figure*}[ht!] 
	\centering
	\makebox[\textwidth][c]{\includegraphics[width=1.0\textwidth]{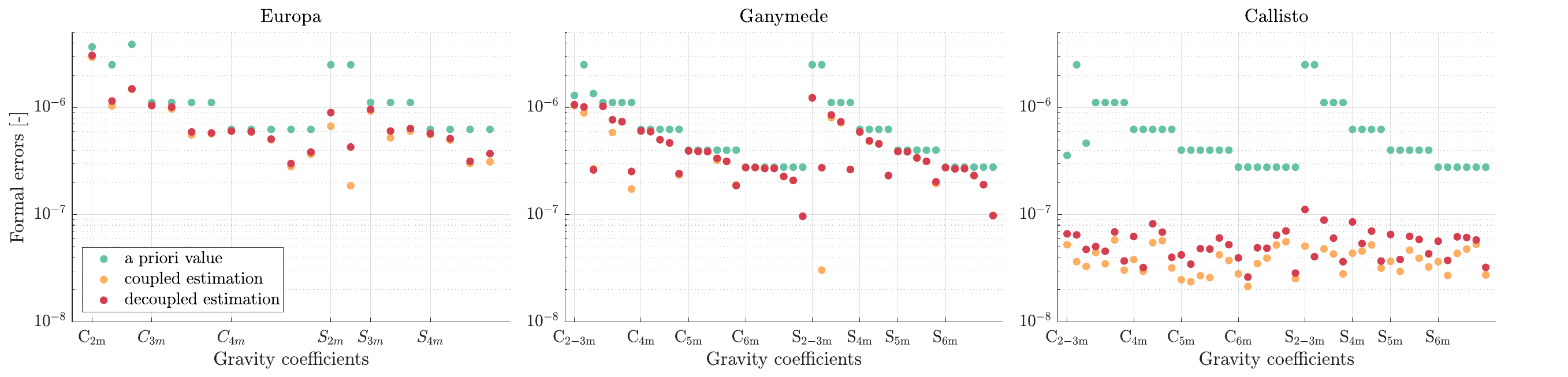}}
	\caption{Formal errors in gravity field coefficients for the three Galilean moons targeted by JUICE flybys. The coupled and decoupled solutions are compared with respect to \textit{a priori} values (see Section \ref{sec:estimationSettings}). \textcolor{black}{The gravity coefficients are plotted along the x-axis: for both the cosine $C_{im}$ and sine $S_{im}$ coefficients, they are grouped by degree and plotted by increasing order $m$. The label $C/S_{im}$ indicates where the coefficients of degree $i$ start, and the order $m$ of these coefficients progressively increases until the start of the next group of coefficients (degree $i+1$).}}
	\label{fig:gravityFieldsFlybys}
\end{figure*}

The formal errors for the moons' gravity field coefficients obtained with the decoupled and coupled approaches are provided in Figure \ref{fig:gravityFieldsFlybys}, superimposed with their \textit{a priori} values (Section \ref{sec:estimationSettings}). \textcolor{black}{The limited number of flybys at Europa (2) and Ganymede (7) does not allow the estimation to significantly improve the gravity field solution with respect to the \textit{a priori} knowledge. This is true for both the coupled and decoupled models, and neither of them seems to perform systematically and distinctly better for these two moons: for the rare coefficients for which an improvement is noticed compared to the \textit{a priori} constraints, the formal errors are sometimes lower with one method, sometimes with the other.} 

Results are different in Callisto's case: the 21 flybys, performed at lower altitudes and condensed over a short period of time (Figure \ref{fig:flybysAltitudes}), help to determine the gravity field well beyond \textit{a priori} values. It is interesting to note that our formal uncertainties for Callisto's gravity field are comparable with those obtained in \cite{diBenedetto2021}. As shown in Figure \ref{fig:gravityFieldsFlybys}, the gravity field uncertainties given by the decoupled solution are slightly larger than those achieved with the coupled model. For some of Callisto's flybys, the decoupled method indeed performs poorly in decorrelating the moon's and spacecraft's arc-wise states, which degrades the gravity field estimation. We must nonetheless stress that these differences remain small, the formal errors obtained with the two estimation strategies still being of comparable orders of magnitude.

The decoupled model was actually used in several gravity field estimation studies, mainly to avoid the challenges arising from the reconstruction of dynamically consistent ephemerides over long timescales, as the dynamical model fidelity could not be brought to the required level \citep[\textit{e.g.}][]{durante2019}. Our results verified that the decoupled errors for the moons' gravity coefficients are also not too optimistic compared to the coupled solution. This analysis thus confirmed that opting for the normal points strategy for the JUICE flyby phase would not notably affect the gravity fields solution. This is particularly relevant for Callisto, while Ganymede's and Europa's gravity field determination will also significantly benefit from JUICE's orbital phase and the Europa Clipper mission, respectively.

\subsubsection{Sensitivity to \textit{a priori} knowledge} \label{sec:sensitivityAPriori}
The decoupled solution was found to strongly depend on the \textit{a priori} constraint applied to the moons' states before each flyby. As an experiment, we discarded the update strategy presented in Section \ref{sec:aPrioriKnowledge} and applied the same \textit{default} \textit{a priori} covariances to all arcs. The normal points approach then led to rather different results, reported in Table \ref{tab:influenceApriori}. All position uncertainties get significantly larger when the state knowledge is not conveyed from one arc to the next. Results are the worst for the moons' radial and tangential positions, with errors increased by more than one order of magnitude. Updating the \textit{a priori} information after each flyby is thus critical if realistic uncertainties are to be achieved with a decoupled approach. In particular, it progressively helps to decorrelate the central moon's and spacecraft's arc-wise dynamics. 

When computing the observations' contribution to the solution using Equation \ref{eq:contributionObservations}, \textcolor{black}{the average $c_\mathbf{q}$ value for the moons' positions drops from $c_\mathbf{q}>$0.98 when using updated \textit{a priori} covariances to $\approx0.40$ with the default ones} (except for the first flyby of each moon for which no updated \textit{a priori} is available and $c_\mathbf{q}$ thus remains close to 1). This confirms that the \textit{a priori} information then becomes predominant and significantly helps the solution.

On the contrary, the coupled solution is not noticeably affected by the adopted \textit{a priori} values for the moons' states ($c_\mathbf{q}\approx 1$), and thus appears significantly more robust. It also relies on a more straightforward, update-free strategy as it only uses the \textit{default a priori} values (see Section \ref{sec:estimationSettings}). 

It must be noted that the \textit{a priori} knowledge for the moons' states, while driving the quality of the decoupled state estimation, has no notable impact on the gravity field solution, irrespective of the selected estimation method. This again shows that the main drawbacks of the normal points strategy do not significantly influence the estimated gravity fields. It confirms that the decoupled method is a good alternative when focusing on gravity field determination, in agreement with conclusions drawn in Section \ref{sec:gravityErrorsFlybyPhase}.

\begin{table}[tbp]
	\caption{Formal errors in position for the Galilean moons in the RTN frame, achieved with the decoupled estimation method for different \textit{a priori} state knowledge. The \textit{a priori} covariances are either updated from one normal point to the next, as described in Section \ref{sec:aPrioriKnowledge}, or kept fixed to their default values for all normal points. The errors were averaged over one year, starting from the first flyby.}
	\label{tab:influenceApriori}
	\centering
	\begin{tabular}{l l r r c}
		\hline
		\textbf{Moons}& & \multicolumn{2}{c}{\textbf{\textit{A priori} values}} & \textbf{Ratio} \\
		\multicolumn{2}{l}{\textbf{(nb flybys)}}& \multicolumn{1}{r}{updated [1]} & constant [2] &  [1]/[2]\\
		\hline
		
		& R &  3.36 km & 35.9 km & 0.09 \\
		Io & T  & 5.36 km & 63.6 km & 0.08 \\
		(0)& N  &  7.15 km&  20.3 km& 0.4 \\ \hline
		
		& R & 0.845 km & 11.5 km& 0.07 \\
		Europa & T  & 1.02 km& 14.2 km& 0.07 \\
		(2)& N &  0.854 km& 9.68 km& 0.09 \\ \hline
		
		& R &  26.5 m& 594 m & 0.04 \\
		Ganymede & T & 26.5 m& 570 m & 0.05 \\
		(7)& N & 57.2 m& 535 m& 0.1 \\ \hline
		
		& R &  3.48 m& 36.1 m & 0.1 \\
		Callisto & T &  4.00 m& 43.9 m& 0.09 \\
		(21)& N& 39.6 m& 118 m & 0.3 \\ \hline
	\end{tabular}
\end{table}

\subsection{Flyby and orbital phases combined} \label{sec:resultsOrbitalPhase}

We extended the tracking data set to include the orbital phase at Ganymede in addition to the flybys, again performing the estimation with both the decoupled and coupled methods. Sections \ref{sec:stateErrorsOrbitalPhase} and \ref{sec:gravityErrorsOrbitalPhase} present the results for the states and gravity field estimates, respectively, obtained in the so-called nominal tracking configuration for the orbital phase (see Section \ref{sec:estimationSettings}). The influence of the tracking settings is further analysed in Section \ref{sec:sensitivityTrackingSchedule}.

\subsubsection{State estimation} \label{sec:stateErrorsOrbitalPhase}

\begin{figure*} [ht!]
	\centering
	\begin{minipage}[l]{0.49\textwidth}
		\centering
		\makebox[\textwidth][c]{\includegraphics[width=1.0\textwidth]{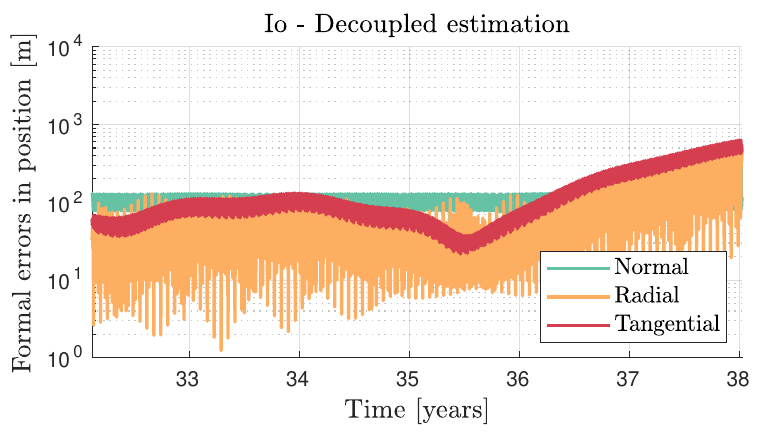}}
		\label{fig:IoDecoupledOrbitalPhase}
	\end{minipage}
	\begin{minipage}[l]{0.49\textwidth}
		\centering
		\makebox[\textwidth][c]{\includegraphics[width=1.0\textwidth]{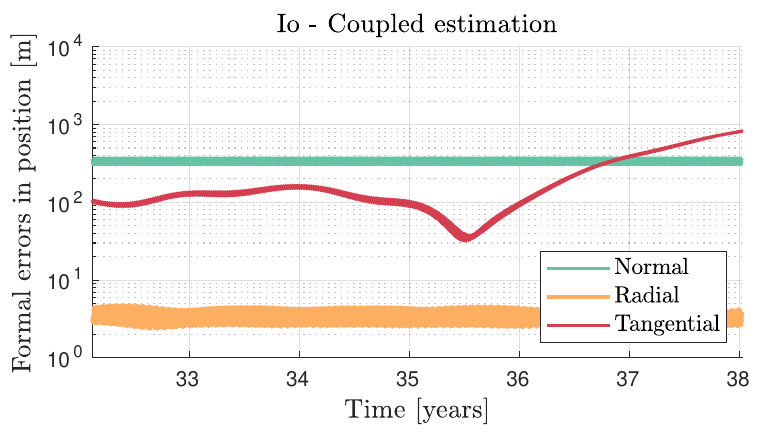}}
		\label{fig:IoCoupledOrbitalPhase}
	\end{minipage}
	\begin{minipage}[l]{0.49\textwidth}
		\centering
		\makebox[\textwidth][c]{\includegraphics[width=1.0\textwidth]{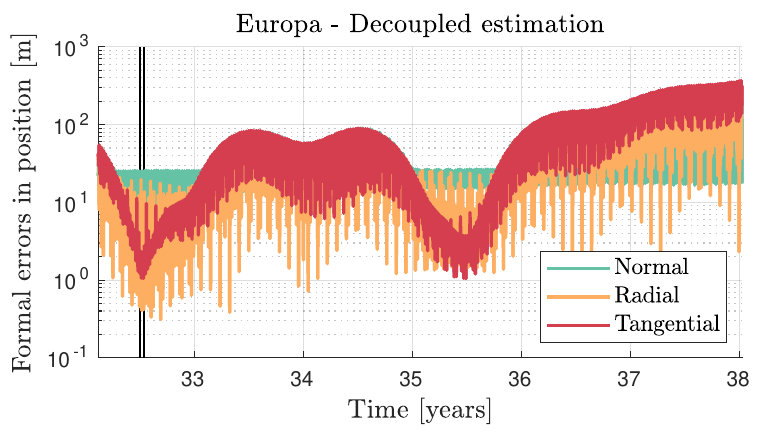}}
		\label{fig:EuropaDecoupledOrbitalPhase}
	\end{minipage}
	\begin{minipage}[l]{0.49\textwidth}
		\centering
		\makebox[\textwidth][c]{\includegraphics[width=1.0\textwidth]{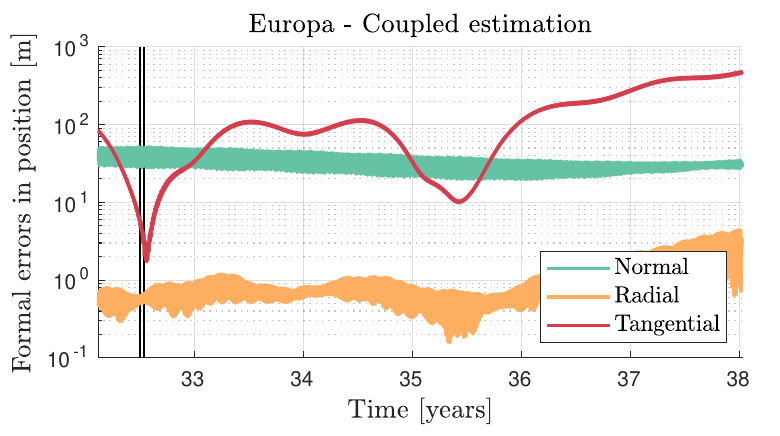}}
		\label{fig:EuropaCoupledOrbitalPhase}
	\end{minipage}
	\begin{minipage}[l]{0.49\textwidth}
		\centering
		\makebox[\textwidth][c]{\includegraphics[width=1.0\textwidth]{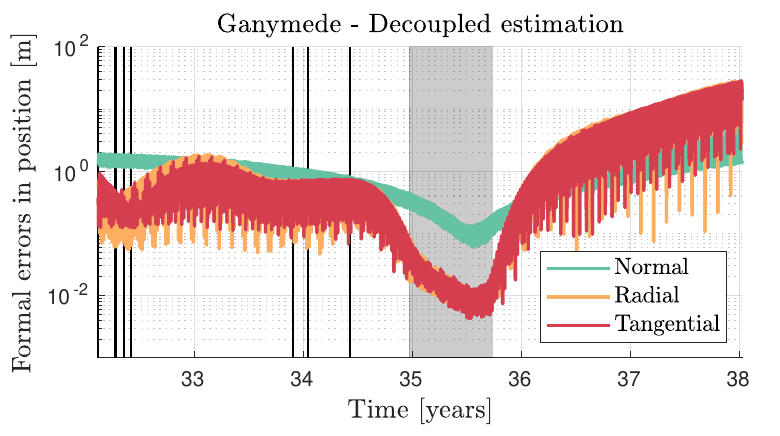}}
		\label{fig:GanymedeDecoupledOrbitalPhase}
	\end{minipage}
	\begin{minipage}[l]{0.49\textwidth}
		\centering
		\makebox[\textwidth][c]{\includegraphics[width=1.0\textwidth]{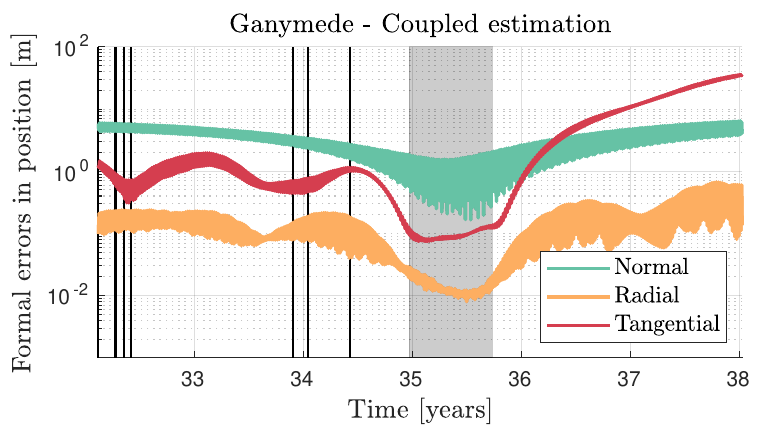}}
		\label{fig:GanymedeCoupledOrbitalPhase}
	\end{minipage}
	\begin{minipage}[l]{0.49\textwidth}
		\centering
		\makebox[\textwidth][c]{\includegraphics[width=1.0\textwidth]{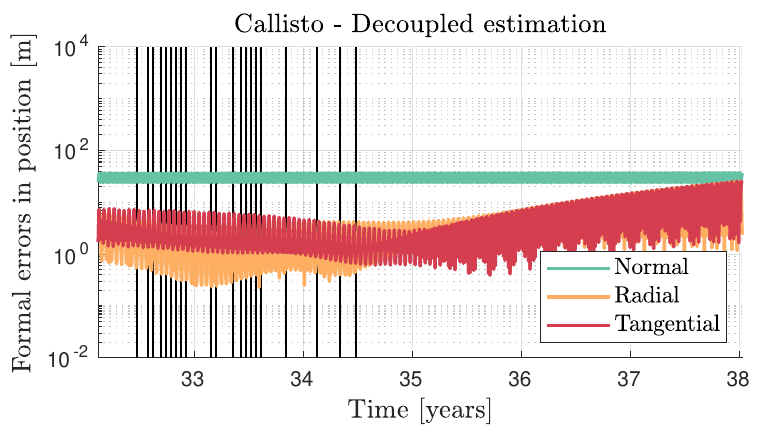}}
		\label{fig:CallistoDecoupledOrbitalPhase}
	\end{minipage}
	\begin{minipage}[l]{0.49\textwidth}
		\centering
		\makebox[\textwidth][c]{\includegraphics[width=1.0\textwidth]{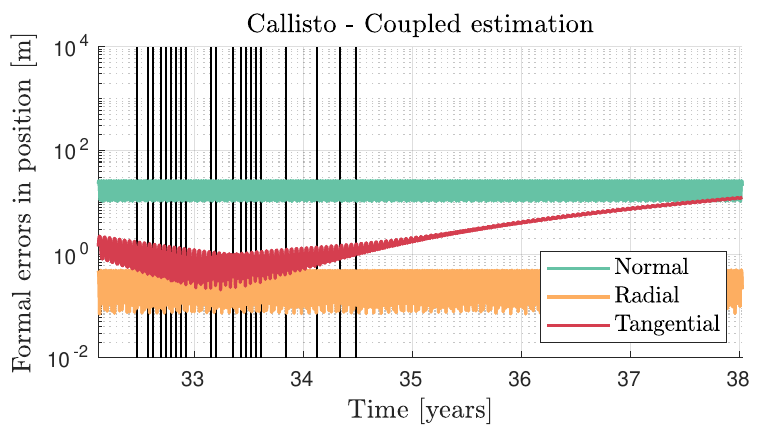}}
		\label{fig:CallistoCoupledOrbitalPhase}
	\end{minipage}
	\caption{\textcolor{black}{Propagated formal errors in position for the four Galilean moons, obtained with the decoupled and coupled state estimation methods (left and right sides, respectively). Tracking data were simulated over both the flybys and orbital phase. The black vertical lines indicate when the JUICE flybys occur, for each moon, and the shaped grey area represents the orbital phase around Ganymede (see Figure \ref{fig:flybysAltitudes}).}}
	\label{fig:stateUncertaintiesOrbitalPhase}
\end{figure*}

To apply the decoupled estimation to the orbital phase, we generated one normal point per tracking arc and determined the central moon's arc-wise state and the associated covariances at the centre of the arc. The propagated uncertainties in the Galilean moons' states are shown in Figure \ref{fig:stateUncertaintiesOrbitalPhase}, for both the decoupled and coupled solutions. Callisto's state being almost only constrained by the flyby phase, the formal errors for this moon are similar to those discussed in Section \ref{sec:resultsFlybyPhase}, when excluding the orbital phase, and are therefore not discussed in the following \textcolor{black}{(see Figures \ref{fig:stateUncertaintiesFlybys} and \ref{fig:stateUncertaintiesOrbitalPhase})}. 

It should be stressed that Ganymede's formal errors fall below 1 m during JUICE orbit, in both the decoupled and coupled cases (Figure \ref{fig:stateUncertaintiesOrbitalPhase}). However, current dynamical models are likely far from being accurate enough to represent sub-meter level effects. Therefore, achieving the presented level of errors in reality will require these models to be rigorously adapted and validated.
The implications of these modelling limitations, which differ for the coupled and decoupled solutions, will be further discussed in Section \ref{sec:discussion}. 

Comparing Figures \ref{fig:stateUncertaintiesFlybys} and \ref{fig:stateUncertaintiesOrbitalPhase} directly highlights the ephemerides improvement provided by the orbital phase. For both the coupled and decoupled solutions, the decrease in Ganymede's position uncertainties during JUICE orbit is clear. These results also clearly illustrate the strong dynamical coupling between the three innermost Galilean moons: the errors reduction for Io and Europa with respect to Figure \ref{sec:stateErrorsFlybyPhase} is indeed achieved by collecting more observations close to Ganymede.

Figure \ref{fig:stateUncertaintiesOrbitalPhase} also confirms the flybys-based conclusions discussed in Section \ref{sec:stateErrorsFlybyPhase}. In particular, the coupled estimation still provides a noticeable improvement in the radial direction, while errors in the moons' tangential positions are overall lower with the decoupled method. It is however interesting to note that these trends are accentuated for some moons, and attenuated for others. For Ganymede, the tangential position uncertainties are noticeably lower with the decoupled model during the orbital phase, and the error reduction in the radial direction achieved by the coupled solution remains limited. The opposite is observed for Io and Europa: the coupled solution's radial position uncertainties are on average one to two orders of magnitude lower than in the decoupled case, while the errors level remains comparable in the tangential direction. 

This is caused by differences in how each method captures the strong dynamical coupling between Io, Europa, and Ganymede. In the decoupled approach, the radiometric data collected during JUICE orbit are used to generate normal points solely at Ganymede. At first, these observations thus exclusively improve our knowledge of Ganymede's local states. The coupling between the three moons is only introduced in the second step of the decoupled estimation (Section \ref{sec:generalPrincipleCoupled}), and Io's and Europa's solutions therefore benefit from the orbital phase in an indirect way, through very accurate normal points generated at Ganymede. On the contrary, the coupled model directly uses all data to estimate the four moons' states concurrently, and provides the most statistically accurate mapping of data uncertainty covariance to parameter covariance (see Section \ref{sec:scopeComparativeAnalysis}).  In the coupled case, the solution improvement provided by the orbital phase is thus more evenly spread between the three innermost moons.  

\subsubsection{Gravity field estimation} \label{sec:gravityErrorsOrbitalPhase}

\begin{figure*}[ht!] 
	\centering
	\makebox[\textwidth][c]{\includegraphics[width=1.0\textwidth]{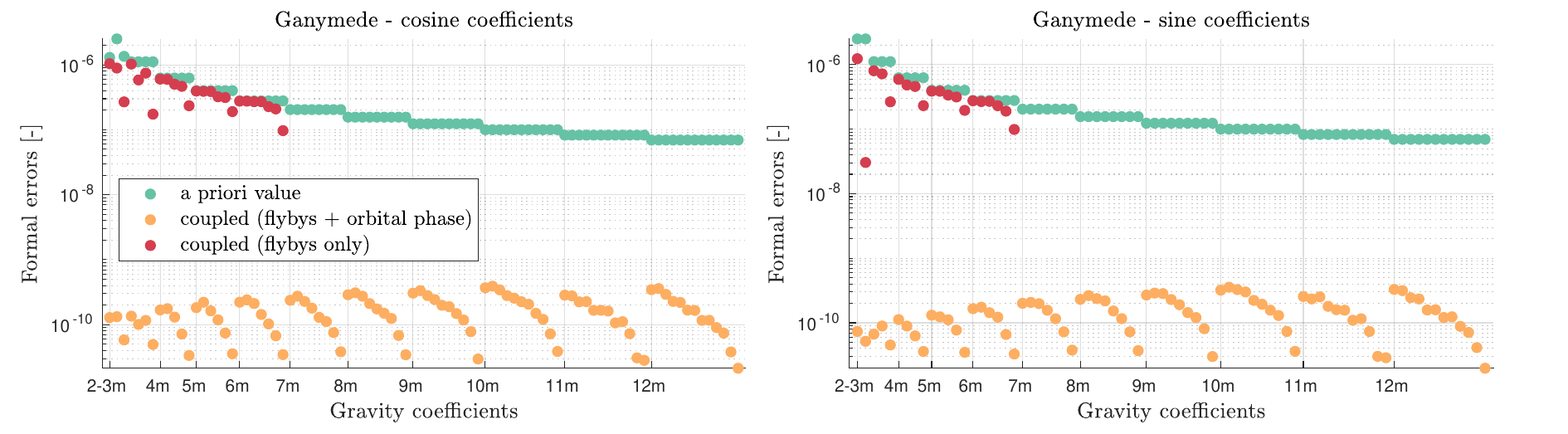}}
	\caption{Formal errors in Ganymede's gravity field coefficients. The coupled solution obtained when including the tracking data collected over both the flybys and orbital phases is compared with the flybys-only solution (up to degree and order 6 only, see Section \ref{sec:estimationSettings}), which was already displayed in Figure \ref{fig:gravityFieldsFlybys}. \textcolor{black}{The gravity coefficients are plotted along the x-axis: they are grouped by degree and plotted by increasing order $m$, with the cosine coefficients first, followed by the sine terms. The label $im$ indicates where the cosine coefficients of degree $i$ start, and the order $m$ of these coefficients progressively increases. They are followed by the sine coefficients of degree $i$ (ordered similarly), until the start of the next group of coefficients (degree $i+1$).}}
	\label{fig:gravityFieldsFullMission}
\end{figure*}

The results and conclusions regarding the moons' estimated gravity fields are similar to those discussed in Section \ref{sec:gravityErrorsFlybyPhase}. For Europa's and Callisto's estimated gravity coefficients, there is actually no noticeable difference with respect to the flybys phase's results, which was expected since the orbital phase at Ganymede does not constrain other moons' gravity fields. The solution for Ganymede is however significantly improved by the orbital phase, as shown in Figure \ref{fig:gravityFieldsFullMission}. It should be noted that these results rely on a simplified estimation setup, and that gravity field studies based on 3GM data from JUICE's orbital phase estimate Ganymede's gravity along with the moon's rotational parameters, Love numbers, \textit{etc.} \citep[\textit{e.g.}][]{cappuccio2020}. Nonetheless, the order of magnitude of the formal uncertainties reported in Figure \ref{fig:gravityFieldsFullMission} are in agreement with those obtained in dedicated 3GM studies \citep{cappuccio2020,diMarchi2021}. 

In our simulations, \textcolor{black}{limited differences between the coupled and decoupled cases could still be detected from the flybys-based results, at least for Callisto whose gravity coefficients could be estimated beyond their \textit{a priori} values (see Figure \ref{fig:gravityFieldsFlybys}).} \textcolor{black}{However, such discrepancies between the two models are not observed anymore (for any moon)} once the orbital phase is included, which is why Figure \ref{fig:gravityFieldsFullMission} only displays the coupled solution. Compared to the flybys, JUICE's \textcolor{black}{orbital phase} generates large amount of data, continuously collected over a longer period of time (as opposed to discrete arcs at each flyby). The contribution of the orbital phase thus completely dominates the gravity field solution (see Figure \ref{fig:gravityFieldsFullMission}). The longer tracking arcs (\textit{i.e.} 2 days, with 8 hours of tracking per day, instead of 8 hours only for the flybys) \textcolor{black}{and the much larger numbers of observations} allow both methods to properly decorrelate the spacecraft's and moon's dynamics, which explains why nearly identical uncertainties are obtained for Ganymede's gravity field in each case. This confirms conclusions drawn in Section \ref{sec:gravityErrorsFlybyPhase}, according to which the adopted state estimation strategy does not significantly influence the gravity field solutions. 

\subsubsection{Sensitivity to tracking settings} \label{sec:sensitivityTrackingSchedule}

We re-ran our simulations with varying arc duration for the orbital phase, to investigate the sensitivity of each estimation method to the tracking configuration. As mentioned in Section \ref{sec:estimationSettings}, three test cases were considered with arcs of one day, two days (nominal) and one week, respectively. When shortening the tracking arcs from two days to one, the errors grow larger for both the decoupled and coupled solutions. The opposite is true when increasing the arc duration to a full week, as longer arcs allow a better decoupling of JUICE's and Ganymede's dynamics. As expected, this effect is the smallest for Callisto's uncertainties. 

\begin{figure*}[ht!] 
	\centering
	\makebox[\textwidth][c]{\includegraphics[width=0.8\textwidth]{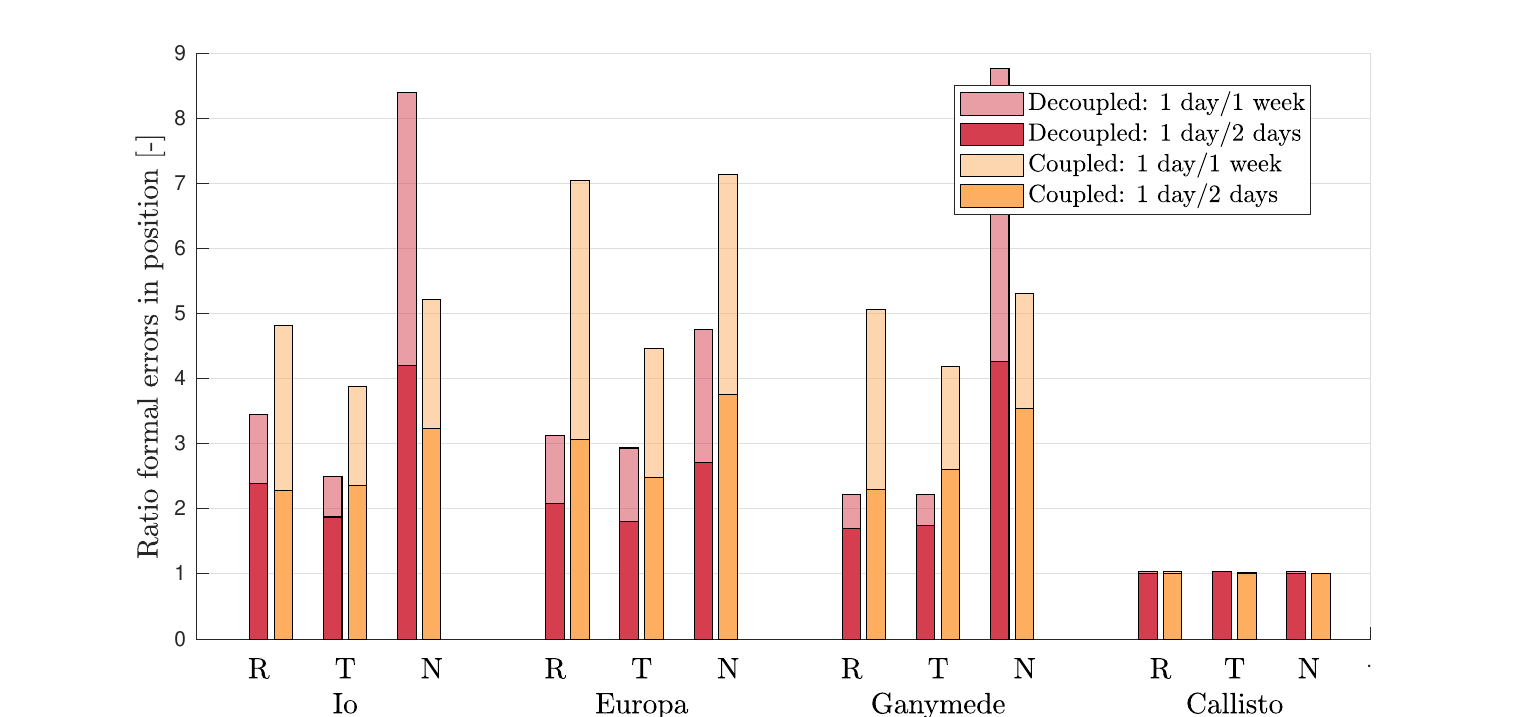}}
	\caption{Ratios of the formal errors in position for different arc durations over JUICE's orbital phase (in the RTN frame: radial, tangential, normal). \textcolor{black}{The results obtained with both the coupled and decoupled methods are displayed for the four Galilean moons (with no noticeable differences for Callisto).} For each method and each moon, we provide the ratio of the formal uncertainties obtained with 1 day-arcs over 2-days arcs, and 1 day-arcs over 1 week-arcs.}
	\label{fig:sensitivityArcDuration}
\end{figure*}

Importantly, the relative evolution of the decoupled and coupled solutions as the arc duration varies provides valuable insights into the fundamental differences between the two state estimation approaches. Figure \ref{fig:sensitivityArcDuration} shows ratios in formal errors resulting from different arc durations. Except for Io's and Ganymede's normal positions, the improvement provided by longer arcs is systematically weaker in the decoupled case.

This is caused by decoupling the spacecraft's and moons' state estimations. As already mentioned, lengthening the tracking arcs generally helps to decorrelate the spacecraft's motion from the central moon's, but does not necessarily reduce the correlations between the moon's own state components. This still results in lower state uncertainties for the normal points. However, it does not automatically reduce the correlations level in the second step of the decoupled estimation, the spacecraft's states being excluded anyway (see Section \ref{sec:estimationSettings}, Table \ref{tab:parameters}). In the decoupled case, the reduction in the moons' state uncertainties thus remains limited by the normal points' high correlations, especially in the radial and tangential directions. On the contrary, the correlations between the spacecraft's and moon's states are directly included in the coupled model, since their dynamics are concurrently estimated. The coupled solution therefore fully benefits from longer tracking arcs, explaining why a more significant improvement is achieved for the radial and tangential positions than what the decoupled model allows.

\section{Discussion: main strengths and challenges of both methods} \label{sec:discussion}

This paper assesses the relative performance of two state estimation strategies, applied to the JUICE test case. To put the obtained formal errors and correlations into context, we discuss practical considerations related to data analysis in the following, and how they affect the coupled and decoupled models in different ways. Data processing challenges are detailed in Section \ref{sec:dataProcessingIssues}, while model-related issues will be addressed in Section \ref{sec:modellingIssues}. As a possible mitigation strategy, we finally suggest a possible alternative state estimation approach in Section \ref{sec:alternativeStrategies}. 

\subsection{Data processing considerations} \label{sec:dataProcessingIssues}

One of the major differences between the coupled and decoupled techniques lies in the way the available data are processed. The decoupled method can theoretically treat each mission, and each mission phase, independently and generate as many normal points as required by the mission design. These normal points can later be combined with those determined from other missions and/or with different observation sets (\textit{e.g.} optical, astrometric data). For the coupled solution, however, all data need to be processed at once.

For the Galilean moons test case in particular, reaching a global solution would ideally require to include spacecraft data from various missions, as well as Earth-based optical astrometric observations. More precisely, JUICE's imbalanced data set with its strong focus on Ganymede would be efficiently complemented by the Europa Clipper and Juno missions, with over 40 flybys at Europa planned for the former \citep[\textit{e.g.}][]{verma2018,tarzi2019}, and an orbital phase at Jupiter lasting since 2016 for the latter, which might be combined with a few crucial flybys at Io during the extended mission phase. In addition to radiometric data, spacecraft-based optical observations captured with navigation cameras can also be useful to help constraining the ephemeris solution. These can be direct imaging of other moons (\textit{e.g.} JANUS data for JUICE, \citealt{dirkx2017}), 
as well as eclipses and Sun or stellar occultations observed from the spacecraft \citep{andreoli2018}.

Additionally, Earth-based photo- and astrometric observations of the Galilean moons have been collected over centuries. They include absolute and differential astrometry, already performed since the 17th century \citep[\textit{e.g.} data starting from late 19th century used in][]{lainey2009}, as well as eclipses and occultations \citep[\textit{e.g.}][]{arlot2019}. More recently, mutual approximations were identified as interesting observables \citep{morgado2019a,fayolle2021} and the first stellar occultation by the moon Europa was observed in 2017, with a remarkable accuracy of 0.80 mas (\textit{i.e.} 2.55 km at Jupiter's distance) \citep{morgado2019b}. Interestingly, the GAIA catalogue will facilitate the observations of such occultations in the future.

Merging all above-mentioned data sets and processing them in a single step does not only make the coupled estimation process slower, but also substantially increases its complexity. It first implies to carefully weigh all different observations to obtain a statistically balanced and realistic solution. The JUICE radiometric data, which led to formal state uncertainties below the meter level for the Galilean moons (see Figure \ref{fig:stateUncertaintiesOrbitalPhase}), would indeed need to be properly combined with Earth-based optical astrometric data whose current accuracy remains larger than 1km (even stellar occultations). Furthermore, optical astrometric and radiometric tracking data are typically processed by different estimation tools, while the coupled estimation model would require a single software able to handle and process all observation types concurrently, which imposes significant practical constraints (many different observation types required, with, for each of them, suitable error and dynamical models, \textit{etc.}).

Moreover, Earth-based observations are collected over longer periods of time compared to the spacecraft data, which are condensed over the planetary missions' timeline. This requires dynamical models to be consistent over both short and long timescales, and represents a major challenge for the reconstruction of a coupled solution, which will be further discussed below.

\subsection{Modelling-related considerations} \label{sec:modellingIssues}

For the JUICE mission specifically, our covariance analysis indicates that Ganymede's formal state uncertainties get lower than the meter level during the orbital phase (see Figure \ref{fig:stateUncertaintiesOrbitalPhase}). For such position errors to be meaningful, major modelling efforts would however be essential. Model-related issues are therefore expected to occur when real JUICE data become available, but their influence on the estimation can unfortunately not be easily quantified in a simulation analysis. \cite{dirkx2016} analysed which effects would likely be detectable from a long-term (several years) signature in the dynamics. However, they did not address the observability and relevance of short-term periodic variations. Such potentially mismodelled dynamical effects (both long- and short-term) are however crucial to discuss here, as they would have different impacts on the coupled and decoupled solutions. 

Among possible sources of inaccuracies, the models currently used for the dissipation occurring inside the moons rely on simplified analytical formulations \citep{lainey2009,lari2018}. These approximations circumvent the need for perfectly consistent rotational and translational models, still unavailable for natural satellites, but are based on time-averaging assumptions. The dissipation inside the moons however directly influences the orbital evolution of the moons themselves and plays a key role in the long-term dynamical history of the Jovian system, such that its accurate modelling is crucial. 

Additionally, due to the coupling between the moons' translational and rotational motions, issues can also originate from mismodelled librations. Ganymede's libration will likely be observable in the dynamics of JUICE itself during the orbital phase \citep{cappuccio2020}. The coupled model thus presents an interesting oppurtunity, since the libration's signature on the spacecraft and moon dynamics is different but would need to be fitted concurrently, possibly yielding a better constrained solution. On the other hand, modelling inconsistencies in Ganymede's librational motion would more critically degrade the residuals of the coupled solution. 

For JUICE specifically, properly modelling the moons' rotations would furthermore require to reconcile what the different instruments are sensitive to. JANUS and GALA (navigation camera and altimeter, respectively) will observe the rotational motion of the moon's surface shell, which might be decoupled from the full body inertial rotation sensed by the radiometric data. Additionally, temporal variations in the central planet's gravity field are also not yet perfectly understood, and are suspected to be responsible for small, unmodelled time-dependent accelerations detected at periapsis in Juno \citep{durante2020} and Cassini data \citep{iess2019}. It should nonetheless be stressed that an imperfect dynamical model can still achieve the required accuracy if properly parametrised (\textit{e.g.} if time-dependent librational and gravitational variations are adequately adjusted), provided the relevant free parameters are incorporated in the estimation, and sufficiently decorrelate from the other parameters.

The fundamental differences between the coupled and decoupled approaches (see Figure \ref{fig:models}) significantly influence how the above-mentioned modelling issues might affect the estimation solutions. Compared to the coupled case, the decoupled approach indeed estimates more state parameters, determined more locally (Table \ref{tab:parameters}), which directly increases the ability of this model to absorb dynamical modelling inaccuracies.

Additionally, the long-term moons' dynamics are only reconstructed in the second step of the decoupled estimation, and they are adjusted to the normal points (\textit{i.e.} arc-wise covariances of the moons' states). For the JUICE flyby phase for example, the decoupled model is thus fitting formal state uncertainties generally ranging from tens up to hundreds of meters (the last flybys at Callisto getting closer to the meter level for the radial and tangential positions). The coupled approach, on the other hand, is directly adjusting the parameters to the radiometric data, with expected accuracies of 20 cm and 15 $\mathrm{\mu}$m/s for the range and range-rate of JUICE with respect to the Earth (Section \ref{sec:estimationSettings}). It is therefore significantly easier to obtain flat residuals with the decoupled estimation strategy (\textit{i.e.} zero-mean residuals, noise within expected observation errors), as observables with accuracies up to $10^1-10^2$ m bring sufficient flexibility to (at least partially) absorb dynamical modelling inaccuracies. 

The fact that the coupled method estimates the moons' dynamics in a single arc also drastically reduces its ability to absorb such model errors. It indeed implies that the estimation model cannot compensate for modelling inaccuracies by a local moon's state variation without affecting other arcs, and possibly conflicting with their own observational constraints. Furthermore, the decoupled estimation leaves out the arc-wise spacecraft states when determining the moons' long-term dynamics (see Section \ref{sec:generalPrincipleDecoupled}). The coupled model, on the contrary, imposes perfect consistency between the spacecraft's and moons' state estimation solutions, which further reduces its degrees of freedom.

In practice, the above entails that the concurrent estimation of the spacecraft's and moons' states would need accurate and consistent dynamical models over both short and long timescales. The current modelling fidelity has for example not yet allowed a coupled solution to be achieved from Cassini and Juno data \citep[\textit{e.g.}][]{durante2019}. Furthermore, even if a global solution can still be reconstructed for the natural bodies' dynamics, modelling errors are expected to manifest themselves in high, incompressible residuals, due to the coupled strategy's lack of flexibility. On the contrary, in the decoupled case, modelling issues are more likely to be absorbed in the final estimated states and to thus remain unnoticed. Reflecting back on the normal points obtained for JUICE, with errors largely below one meter, more realistic uncertainties in agreement with the available dynamical models would significantly raise this error level. It would therefore degrade the quality of the decoupled estimation, but not prevent the obtention of a viable solution. On the other hand, the (large) post-fit residuals obtained with the coupled estimation can be indicative of the magnitude of the dynamical modelling inaccuracies, and help to interpret the decoupled solution's true uncertainty. 

\subsection{Possible alternative strategy} \label{sec:alternativeStrategies}

Despite the promising results obtained with the coupled model, Sections \ref{sec:dataProcessingIssues} and \ref{sec:modellingIssues} highlighted crucial challenges that would need to be addressed before a reliable and statistically consistent coupled solution can be reconstructed for the Galilean moons' dynamics. Hybrid approaches, halfway between fully coupled and decoupled strategies, could therefore prove promising. In such hybrid scenarios, the coupled model can be applied more locally, rather than on the entire time period of interest. For the JUICE test case, a coupled solution might typically be attainable over the GEO/GCO5000 and/or GCO500 phases. Some flybys could also be processed concurrently (\textit{e.g.} the two flybys at Europa). It would also be possible to reconstruct a coupled solution over the entire JUICE mission, but to combine it with other mission/data sets in a separate step, to mitigate the data merging issues highlighted in Section \ref{sec:dataProcessingIssues}.

This would efficiently mitigate the effects of long-term modelling inconsistencies and thus make a coupled solution achievable \textit{locally}. Such \textit{local} coupled solutions can then be treated as normal points or a priori information for other analyses, and combined with those generated for different missions or with optical astrometric data, as discussed in Section \ref{sec:dataProcessingIssues}. This would also eliminate the need to process astrometric and radiometric data in a single estimation tool as in the fully coupled strategy (see Section \ref{sec:dataProcessingIssues}), by splitting the data analysis steps. This hybrid approach could represent an interesting alternative: it would potentially allow the estimation solution to benefit from a higher sensitivity to the system's full dynamical coupling, while still guaranteeing that a viable solution can be achieved.

\section{Conclusions} \label{sec:conclusions}

We provided the complete formulation for a coupled, concurrent state estimation of the spacecraft and natural bodies from planetary missions' data. Such a coupled model has already been used in past studies \citep[\textit{i.e.}][]{dirkx2018,lari2019}, but, to the best of our knowledge, was not explicitly described in the literature. We then performed a detailed covariance analysis comparing the decoupled and coupled estimation strategies for the upcoming JUICE mission. The realism of the formal errors given by the coupled model was verified by running a deterministic simulation and comparing the least-squares estimation errors with the formal uncertainties.

The JUICE mission will make us face both unique opportunities and challenges due to the unusual mission profile and unprecedented accuracy of the radiometric data, used to reconstruct the strongly coupled dynamics of the Galilean moons. Our study primarily assessed how a coupled solution, if attainable, would affect the accuracy of the Galilean moons' ephemerides. \textcolor{black}{The results of the decoupled estimation approach, on the other hand, indicate the uncertainty level that would be achievable in case the coupled model failed to  reconstruct a viable solution for the moons' dynamics (see discussion in Section \ref{sec:discussion}).} It must be stressed that we do not anticipate the conclusions of our study to depend on the CReMA (\textit{i.e.} trajectory) version used for the JUICE spacecraft. While the absolute uncertainty levels for the estimated parameters might vary, the comparison between the two state estimation methods is expected to yield similar results.

We first showed that selecting appropriate \textit{a priori} values for the moons' states is critical for the decoupled solution. The state solution improvement must be accounted for by updating the \textit{a priori} covariance from one normal point to the next. Discarding the information gained in previous arcs indeed leads to poorly constrained normal points (limited to kilometer level accuracy). This eventually drives the decoupled solution to be one or two orders of magnitude less accurate than the coupled one, for all moons and in all directions (\textit{e.g.} tens of meters against $\sim500$ m for Ganymede's position uncertainties with and without \textit{a priori} update, respectively, see Table \ref{tab:influenceApriori}). 

Furthermore, the flybys-based results already highlighted notable differences between the coupled and decoupled state solutions. The coupled method achieves lower radial position uncertainties ($\sim10^2$ m for Io and Europa, $\sim10^1$ m for Ganymede, $\sim10^0$ m for Callisto), which are one or two orders of magnitude smaller than in the decoupled case. In the tangential direction, the formal errors given by the decoupled approach are however slightly lower: $\sim10$ m against $\sim30$ m, keeping Ganymede's initial position uncertainties as an example. As discussed in Section \ref{sec:scopeComparativeAnalysis}, the coupled model is nonetheless considered to provide a statistically accurate representation of the estimation problem, which might imply that the tangential position uncertainties obtained with the decoupled model are slightly optimistic. 

\textcolor{black}{Our results also proved that JUICE orbital phase is crucial for the quality of the ephemerides of the four Galilean moons.} Assuming a perfect dynamical model, Ganymede's position uncertainties would even reach submetric levels during the orbit. Interestingly, it seems that Io's and Europa's solutions benefit more from the orbital phase when adopting a coupled approach (Europa's radial position also determined with an accuracy better than one meter, see Figure \ref{sec:stateErrorsOrbitalPhase}), while the improvement appears stronger for Ganymede in the decoupled case. Including the orbital phase in our simulations actually enhances the differences between the two estimation strategies, and indicates that the dynamical coupling between the Galilean moons is better captured by the coupled model. 

Concerning dynamical parameters, the adopted state estimation method is not so influential for the gravity field coefficients' estimates. This indicates that the normal points strategy is perfectly adapted for gravity field determination from JUICE radiometric data \citep[\textit{e.g.}][]{magnanini2021}, and generally for most local parameters studies as well. \textcolor{black}{Rotational and tidal dissipation parameters}, on the other hand, were omitted from our simulated estimations. Future studies should include these parameters and try determining the dissipation in Jupiter at the moons' frequency \citep{lainey2009,lainey2020}. It would then be crucial to investigate whether the decoupled model can achieve realistic uncertainties despite a slightly optimistic estimation of the moons' along-track motion.

As a preliminary step towards an improved solution for the Galilean moons' ephemerides, we limited ourselves to an analysis based on simulated data only. As extensively discussed in Section \ref{sec:discussion}, many issues are however expected to arise when real JUICE radiometric data become available and are fed to the estimation process, as happened with Cassini and Juno data \citep{durante2019,durante2020}. For the JUICE test case, the coupled estimation model nonetheless appears to yield promising improvements with respect to the decoupled strategy: it overall reconstructs a more balanced solution for the three Galilean moons in resonance, with significantly reduced uncertainties in the moons' radial positions. Despite the challenging issues that may emerge, this motivates future efforts to achieve a coupled state estimation for the Galilean moons. 

\section*{Acknowledgments}

The authors would like to thank the ZENITH working group for the many fruitful discussions. In particular, insights from M. Zannoni, L. Gomez-Casajus and A. Magnanini were extremely valuable. This research was partially funded by ESA's OSIP (Open Space Innovation Platform) program. This work was supported by CNES, focused on JUICE.

\bibliographystyle{apalike} 

\bibliography{references}

\begin{thebibliography}{}

\bibitem[Acton~Jr, 1996]{acton1996}
Acton~Jr, C.~H. (1996).
\newblock {Ancillary data services of NASA's Navigation and Ancillary
  Information Facility}.
\newblock {\em Planetary and Space Science}, 44(1):65--70.

\bibitem[Alessi et~al., 2012]{alessi2012}
Alessi, E.~M., Cicalo, S., Milani, A., and Tommei, G. (2012).
\newblock {Desaturation manoeuvres and precise orbit determination for the
  BepiColombo mission}.
\newblock {\em Monthly Notices of the Royal Astronomical Society},
  423(3):2270--2278.

\bibitem[Andreoli and Zannoni, 2018]{andreoli2018}
Andreoli, F. and Zannoni, M. (2018).
\newblock {Improvement of Jupiter’s satellites ephemerides using stellar
  occultation observations}.
\newblock Master's thesis, University of Bologna.

\bibitem[Arlot and Emelyanov, 2019]{arlot2019}
Arlot, J.-E. and Emelyanov, N. (2019).
\newblock {Natural satellites mutual phenomena observations: achievements and
  future}.
\newblock {\em Planetary and Space Science}, 169:70--77.

\bibitem[Bocanegra-Baham{\'o}n et~al., 2018]{bocanegra2018}
Bocanegra-Baham{\'o}n, T., Calv{\'e}s, G.~M., Gurvits, L., Duev, D.,
  Pogrebenko, S., Cim{\`o}, G., Dirkx, D., and Rosenblatt, P. (2018).
\newblock {Planetary radio interferometry and Doppler experiment (PRIDE)
  technique: a test case of the mars express phobos flyby-II. Doppler tracking:
  formulation of observed and computed values, and noise budget}.
\newblock {\em Astronomy \& Astrophysics}, 609:A59.

\bibitem[Cappuccio et~al., 2020]{cappuccio2020}
Cappuccio, P., Hickey, A., Durante, D., Di~Benedetto, M., Iess, L., De~Marchi,
  F., Plainaki, C., Milillo, A., and Mura, A. (2020).
\newblock {Ganymede's gravity, tides and rotational state from JUICE's 3GM
  experiment simulation}.
\newblock {\em Planetary and Space Science}, 187:104902.

\bibitem[Charlot et~al., 2020]{charlot2020}
Charlot, P., Jacobs, C., Gordon, D., Lambert, S., de~Witt, A., B{\"o}hm, J.,
  Fey, A., Heinkelmann, R., Skurikhina, E., Titov, O., et~al. (2020).
\newblock {The third realization of the International Celestial Reference Frame
  by very long baseline interferometry}.
\newblock {\em Astronomy \& Astrophysics}, 644:A159.

\bibitem[De~Marchi et~al., 2021]{diMarchi2021}
De~Marchi, F., Di~Achille, G., Mitri, G., Cappuccio, P., Di~Stefano, I.,
  Di~Benedetto, M., and Iess, L. (2021).
\newblock {Observability of Ganymede's gravity anomalies related to surface
  features by the 3GM experiment onboard ESA's JUpiter ICy moons Explorer
  (JUICE) mission}.
\newblock {\em Icarus}, 354:114003.

\bibitem[Di~Benedetto et~al., 2021]{diBenedetto2021}
Di~Benedetto, M., Cappuccio, P., Molli, S., Federici, L., and Zavoli, A.
  (2021).
\newblock {Analysis of 3GM Callisto Gravity Experiment of the JUICE Mission}.
\newblock {\em arXiv preprint arXiv:2101.03401}.

\bibitem[Di~Ruscio, 2021]{diruscio2021}
Di~Ruscio, A. (2021).
\newblock {\em {Use of radio science and navigation data for the construction
  of planetary ephemerides}}.
\newblock PhD thesis, Universit{\'e} C{\^o}te d'Azur; Universit{\`a} degli
  studi La Sapienza (Rome).

\bibitem[Di~Ruscio et~al., 2020]{diruscio2020}
Di~Ruscio, A., Fienga, A., Durante, D., Iess, L., Laskar, J., and Gastineau, M.
  (2020).
\newblock {Analysis of Cassini radio tracking data for the construction of
  INPOP19a: A new estimate of the Kuiper belt mass}.
\newblock {\em Astronomy \& Astrophysics}, 640:A7.

\bibitem[Dirkx et~al., 2017]{dirkx2017}
Dirkx, D., Gurvits, L.~I., Lainey, V., Lari, G., Milani, A., Cim{\`o}, G.,
  Bocanegra-Bahamon, T., and Visser, P. (2017).
\newblock {On the contribution of PRIDE-JUICE to Jovian system ephemerides}.
\newblock {\em Planetary and Space Science}, 147:14--27.

\bibitem[Dirkx et~al., 2016]{dirkx2016}
Dirkx, D., Lainey, V., and Gurvits, L. ans~Visser, P. (2016).
\newblock {{Dynamical Modelling of the Galilean Moons for the JUICE Mission}}.
\newblock {\em {Planetary and Space Science}}, 134:82--95.

\bibitem[Dirkx et~al., 2019]{dirkx2019}
Dirkx, D., Mooij, E., and Root, B. (2019).
\newblock {Propagation and estimation of the dynamical behaviour of
  gravitationally interacting rigid bodies}.
\newblock {\em Astrophysics and Space Science}, 364(2):37.

\bibitem[Dirkx et~al., 2018]{dirkx2018}
Dirkx, D., Prochazka, I., Bauer, S., Visser, P., Noomen, R., Gurvits, L.~I.,
  and Vermeersen, B. (2018).
\newblock {Laser and radio tracking for planetary science missions—a
  comparison}.
\newblock {\em Journal of Geodesy}, pages 1--16.

\bibitem[Duev et~al., 2016]{duev2016}
Duev, D., Pogrebenko, S., Cimò, G., Calvés, G.~M., Bahamón, T.~B., Kettenis,
  L. G.~M., Kania, J., Tudose, V., Rosenblatt, P., Marty, J.-C., Lainey, V.,
  de~Vicente, P., Quick, J., Nicola, M., Neidhard, A., Kronschnabl, G., Ploetz,
  G., Haas, R., Lindquist, M., Orlatti, A., Ipatov, A., Kharinov, M.,
  Mikhailov, A., Gulyaev, S., Weston, S., Natush, T., Zhang, W., Wang, W., Bo,
  X., Yang, W., Hao, L., and Kallunki, J. (2016).
\newblock {{Planetary Radio Interferometry and Doppler Experiment (PRIDE)
  technique: a test case of the Mars Express Phobos fly-by}}.
\newblock {\em {Astronomy and Astrophysics}}, 593:A34.

\bibitem[{Duev} et~al., 2012]{duev2012}
{Duev}, D.~A., {Molera Calv{\'e}s}, G., {Pogrebenko}, S.~V., {Gurvits}, L.~I.,
  {Cim{\'o}}, G., and {Bocanegra Bahamon}, T. (2012).
\newblock {{Spacecraft VLBI and Doppler tracking: algorithms and
  implementation}}.
\newblock {\em Astronomy and Astrophysics}, 541:A43.

\bibitem[Durante et~al., 2019]{durante2019}
Durante, D., Hemingway, D., Racioppa, P., Iess, L., and Stevenson, D. (2019).
\newblock {{Titan's gravity field and interior structure after Cassini}}.
\newblock {\em Icarus}, 326:123--132.

\bibitem[Durante et~al., 2020]{durante2020}
Durante, D., Parisi, M., Serra, D., Zannoni, M., Notaro, V., Racioppa, P.,
  Buccino, D., Lari, G., Gomez~Casajus, L., Iess, L., et~al. (2020).
\newblock {{Jupiter's gravity field halfway through the Juno mission}}.
\newblock {\em Geophysical Research Letters}, 47(4):e2019GL086572.

\bibitem[Fayolle et~al., 2021]{fayolle2021}
Fayolle, M., Dirkx, D., Visser, P., and Lainey, V. (2021).
\newblock {Analytical framework for mutual approximations-Derivation and
  application to Jovian satellites}.
\newblock {\em Astronomy \& Astrophysics}, 652:A93.

\bibitem[Floberghagen, 2001]{floberghagen2001}
Floberghagen (2001).
\newblock {\em {{The Far Side: Lunar Gravimetry into the Third Millenium}}}.
\newblock PhD thesis, Delft University of Technology.

\bibitem[Fuller et~al., 2016]{fuller2016}
Fuller, J., Luan, J., and Quataert, E. (2016).
\newblock {Resonance locking as the source of rapid tidal migration in the
  Jupiter and Saturn moon systems}.
\newblock {\em Monthly Notices of the Royal Astronomical Society},
  458(4):3867--3879.

\bibitem[Genova et~al., 2021]{genova2021}
Genova, A., Hussmann, H., Van~Hoolst, T., Heyner, D., Iess, L., Santoli, F.,
  Thomas, N., Cappuccio, P., di~Stefano, I., Kolhey, P., et~al. (2021).
\newblock {Geodesy, geophysics and fundamental physics investigations of the
  BepiColombo mission}.
\newblock {\em Space science reviews}, 217(2):1--62.

\bibitem[{Greenberg}, 2010]{greenberg2010}
{Greenberg}, R. (2010).
\newblock {{The icy Jovian satellites after the Galileo mission}}.
\newblock {\em Reports on Progress in Physics}, 73(3):036801.

\bibitem[{Gurvits} et~al., 2013]{gurvits2013}
{Gurvits}, L.~I., {Bocanegra Bahamon}, T.~M., {Cim{\`o}}, G., {Duev}, D.~A.,
  {Molera Calv{\'e}s}, G., {Pogrebenko}, S.~V., {de Pater}, I., {Vermeersen},
  L.~L.~A., {Rosenblatt}, P., {Oberst}, J., {Charlot}, P., {Frey}, S., and
  {Tudose}, V. (2013).
\newblock {{Planetary Radio Interferometry and Doppler Experiment (PRIDE) for
  the JUICE mission}}.
\newblock {\em European Planetary Science Congress 2013}, 8:357.

\bibitem[Hay et~al., 2020]{hay2020}
Hay, H.~C., Trinh, A., and Matsuyama, I. (2020).
\newblock {Powering the Galilean Satellites with Moon-Moon Tides}.
\newblock {\em Geophysical Research Letters}, 47(15):e2020GL088317.

\bibitem[Hussmann and Spohn, 2004]{hussmannSpohn2004}
Hussmann, H. and Spohn, T. (2004).
\newblock {Thermal-orbital evolution of Io and Europa}.
\newblock {\em Icarus}, 171(2):391--410.

\bibitem[Iess et~al., 2018]{iess2018}
Iess, L., Folkner, W., Durante, D., Parisi, M., Kaspi, Y., Galanti, E.,
  Guillot, T., Hubbard, W., Stevenson, D., Anderson, J., et~al. (2018).
\newblock {Measurement of Jupiter’s asymmetric gravity field}.
\newblock {\em Nature}, 555(7695):220.

\bibitem[Iess et~al., 2019]{iess2019}
Iess, L., Militzer, B., Kaspi, Y., Nicholson, P., Durante, D., Racioppa, P.,
  Anabtawi, A., Galanti, E., Hubbard, W., Mariani, M., et~al. (2019).
\newblock {Measurement and implications of Saturn’s gravity field and ring
  mass}.
\newblock {\em Science}, 364(6445).

\bibitem[Jones et~al., 2020]{jones2020}
Jones, D.~L., Folkner, W.~M., Jacobson, R.~A., Jacobs, C.~S., Romney, J., and
  Dhawan, V. (2020).
\newblock {Very Long Baseline Array Astrometry of Cassini: The Final Epochs and
  an Improved Orbit of Saturn}.
\newblock {\em The Astronomical Journal}, 159(2):72.

\bibitem[Jones et~al., 2015]{jones2015}
Jones, D.~L., Folkner, W.~M., Jacobson, R.~A., Jacobs, C.~S., Romney, J.,
  Dhawan, V., and Fomalont, E.~B. (2015).
\newblock {Update on VLBA Astrometry of Cassini}.
\newblock In {\em American Astronomical Society Meeting Abstracts\# 225},
  volume 225, pages 137--12.

\bibitem[Kaula, 1966]{kaula1966}
Kaula, W.~M. (1966).
\newblock {Tesseral harmonics of the Earth's gravitational field from camera
  tracking of satellites}.
\newblock {\em Journal of Geophysical research}, 71(18):4377--4388.

\bibitem[{Konopliv} et~al., 2011]{konopliv2011}
{Konopliv}, A.~S., {Asmar}, S.~W., {Folkner}, W.~M., {Karatekin}, {\"O}.,
  {Nunes}, D.~C., {Smrekar}, S.~E., {Yoder}, C.~F., and {Zuber}, M.~T. (2011).
\newblock {{Mars high resolution gravity fields from MRO, Mars seasonal
  gravity, and other dynamical parameters}}.
\newblock {\em {Icarus}}, 211:401--428.

\bibitem[Lainey et~al., 2009]{lainey2009}
Lainey, V., Arlot, J.-E., Karatekin, {\"O}., and Van~Hoolst, T. (2009).
\newblock {Strong tidal dissipation in Io and Jupiter from astrometric
  observations}.
\newblock {\em Nature}, 459(7249):957--959.

\bibitem[Lainey et~al., 2020]{lainey2020}
Lainey, V., Casajus, L.~G., Fuller, J., Zannoni, M., Tortora, P., Cooper, N.,
  Murray, C., Modenini, D., Park, R.~S., Robert, V., et~al. (2020).
\newblock {{Resonance locking in giant planets indicated by the rapid orbital
  expansion of Titan}}.
\newblock {\em Nature Astronomy}, pages 1--6.

\bibitem[{Lainey} et~al., 2004]{lainey2004}
{Lainey}, V., {Duriez}, L., and {Vienne}, A. (2004).
\newblock {{New accurate ephemerides for the Galilean satellites of Jupiter. I.
  Numerical integration of elaborated equations of motion}}.
\newblock {\em {Astronomy and Astrophysics}}, 420:1171--1183.

\bibitem[{Lainey} and {Tobie}, 2005]{laineyTobie2005}
{Lainey}, V. and {Tobie}, G. (2005).
\newblock {{New constraints on Io's and Jupiter's tidal dissipation}}.
\newblock {\em Icarus}, 179:485--489.

\bibitem[Lari, 2018]{lari2018}
Lari, G. (2018).
\newblock {A semi-analytical model of the Galilean satellites’ dynamics}.
\newblock {\em Celestial Mechanics and Dynamical Astronomy}, 130(8):1--25.

\bibitem[Lari and Milani, 2018]{lari2018b}
Lari, G. and Milani, A. (2018).
\newblock {The Galilean satellites’ dynamics and the estimation of the Jovian
  system’s dissipation from JUICE data}.
\newblock {\em Univerit{\'a} di Pisa}.

\bibitem[Lari and Milani, 2019]{lari2019}
Lari, G. and Milani, A. (2019).
\newblock {Chaotic orbit determination in the context of the JUICE mission}.
\newblock {\em Planetary and Space Science}, 176:104679.

\bibitem[Lunine, 2017]{lunine2017}
Lunine, J.~I. (2017).
\newblock {{Ocean worlds exploration}}.
\newblock {\em Acta Astronautica}, 131:123--130.

\bibitem[Magnanini, 2021]{magnanini2021}
Magnanini, A. (2021).
\newblock {Estimation of the Ephemerides and Gravity Fields of the Galilean
  Moons Through Orbit Determination of the JUICE Mission}.
\newblock {\em Aerotecnica Missili \& Spazio}, 100(3):195--206.

\bibitem[Mazarico et~al., 2015]{mazarico2015}
Mazarico, E., Genova, A., Neumann, G.~A., Smith, D.~E., and Zuber, M.~T.
  (2015).
\newblock {Simulated recovery of Europa's global shape and tidal Love numbers
  from altimetry and radio tracking during a dedicated flyby tour}.
\newblock {\em Geophysical Research Letters}, 42(9):3166--3173.

\bibitem[Mazarico et~al., 2017]{mazarico2017}
Mazarico, E., Rowlands, D.~D., Sabaka, T.~J., Getzandanner, K.~M., Rubincam,
  D.~P., Nicholas, J.~B., and Moreau, M.~C. (2017).
\newblock {Recovery of Bennu’s orientation for the OSIRIS-REx mission:
  implications for the spin state accuracy and geolocation errors}.
\newblock {\em Journal of Geodesy}, 91(10):1141--1161.

\bibitem[Milani and Gronchi, 2010]{milaniGronchi2010}
Milani, A. and Gronchi, G. (2010).
\newblock {\em {Theory of Orbit Determination}}.
\newblock Theory of Orbit Determination. Cambridge University Press.

\bibitem[Molera~Calv{\'e}s et~al., 2021]{molera2021}
Molera~Calv{\'e}s, G., Pogrebenko, S.~V., Wagner, J.~F., Cim{\`o}, G., Gurvits,
  L.~I., Bocanegra-Baham{\'o}n, T.~M., Duev, D.~A., and Nunes, N.~V. (2021).
\newblock {High spectral resolution multi-tone Spacecraft Doppler tracking
  software: Algorithms and implementations}.
\newblock {\em Publications of the Astronomical Society of Australia}, 38.

\bibitem[Montenbruck and Gill, 2000]{montenbruckGill2000}
Montenbruck, O. and Gill, E. (2000).
\newblock {\em {Satellite Orbits: Models, Methods, and Applications}}.
\newblock Physics and Astronomy Online Library. Springer Verlag.

\bibitem[Morgado et~al., 2019a]{morgado2019b}
Morgado, B., Benedetti-Rossi, G., Gomes-J{\'u}nior, A., Assafin, M., Lainey,
  V., Vieira-Martins, R., Camargo, J., Braga-Ribas, F., Boufleur, R., Fabrega,
  J., et~al. (2019a).
\newblock {First stellar occultation by the Galilean moon Europa and upcoming
  events between 2019 and 2021}.
\newblock {\em Astronomy \& Astrophysics}, 626:L4.

\bibitem[Morgado et~al., 2019b]{morgado2019a}
Morgado, B., Vieira-Martins, R., Assafin, M., Machado, D., Camargo, J., Sfair,
  R., Malacarne, M., Braga-Ribas, F., Robert, V., Bassallo, T., et~al. (2019b).
\newblock {APPROX--mutual approximations between the Galilean moons: the
  2016--2018 observational campaign}.
\newblock {\em Monthly Notices of the Royal Astronomical Society},
  482(4):5190--5200.

\bibitem[{Peale}, 1999]{peale1999}
{Peale}, S.~J. (1999).
\newblock {{Origin and Evolution of the Natural Satellites}}.
\newblock {\em Annual Review of Astronomy and Astrophysics}, 37:533--602.

\bibitem[{Pogrebenko} et~al., 2004]{pogrebenko2004}
{Pogrebenko}, S.~V., {Gurvits}, L.~I., {Campbell}, R.~M., {Avruch}, I.~M.,
  {Lebreton}, J.-P., and {van't Klooster}, C.~G.~M. (2004).
\newblock {{VLBI tracking of the Huygens probe in the atmosphere of Titan}}.
\newblock In {A.~Wilson}, editor, {\em Planetary Probe Atmospheric Entry and
  Descent Trajectory Analysis and Science}, volume 544 of {\em ESA Special
  Publication}, pages 197--204.

\bibitem[Rosenblatt et~al., 2008]{rosenblatt2008}
Rosenblatt, P., Lainey, V., Le~Maistre, S., Marty, J., Dehant, V., P{\"a}tzold,
  M., Van~Hoolst, T., and H{\"a}usler, B. (2008).
\newblock {Accurate Mars Express orbits to improve the determination of the
  mass and ephemeris of the Martian moons}.
\newblock {\em Planetary and Space science}, 56(7):1043--1053.

\bibitem[{Schubert} et~al., 2004]{schubert2004}
{Schubert}, G., {Anderson}, J.~D., {Spohn}, T., and {McKinnon}, W.~B. (2004).
\newblock {{Interior composition, structure and dynamics of the Galilean
  satellites}}.
\newblock In {Bagenal}, F., {Dowling}, T.~E., and {McKinnon}, W.~B., editors,
  {\em Jupiter.~The Planet, Satellites and Magnetosphere}, pages 281--306.

\bibitem[Serra et~al., 2018]{serra2018}
Serra, D., Spoto, F., and Milani, A. (2018).
\newblock {A multi-arc approach for chaotic orbit determination problems}.
\newblock {\em Celestial Mechanics and Dynamical Astronomy}, 130(11):1--17.

\bibitem[Tarzi et~al., 2019]{tarzi2019}
Tarzi, Z., Boone, D., Mastrodemos, N., Nandi, S., and Young, B. (2019).
\newblock {Orbit determination sensitivity analysis for the Europa Clipper
  Mission tour}.

\bibitem[Verma and Margot, 2018]{verma2018}
Verma, A.~K. and Margot, J.-L. (2018).
\newblock {Expected precision of Europa Clipper gravity measurements}.
\newblock {\em Icarus}, 314:35--49.

\end{thebibliography}

\appendix

\setcounter{figure}{0}
\setcounter{table}{0}

\section{\textcolor{black}{Influence of the non-central moons' states as consider parameters on the normal points determination}} \label{appendix:nonCentralMoonsConsiderParameters}

\textcolor{black}{This appendix presents a subset of the results verifying how the normal point generated for a flyby's central moon is affect by uncertainties in the other moons' states. As mentioned in Section \ref{sec:generalPrincipleDecoupled}, the first step of the decoupled model only estimates the arc-wise state solution for the central moon and neglects the possible contribution of these uncertainties. As a verification, we quantified the influence of this simplification by adding the non-central moons' states as consider parameters for each normal point determination.}

\textcolor{black}{We only show the results obtained for two specific flybys. One is the first flyby of JUICE's flyby tour, which is performed around Ganymede. Since the uncertainties in the other moons' states are then still limited to their \textit{a priori} knowledge, we could expect their influence on Ganymede's estimated state solution to be significant. The 12th flyby at Callisto was also selected as a particularly interesting one: the uncertainties of Callisto's normal points are indeed noticeably reduced from this flyby onward (formal errors smaller than 10 m in the radial and tangential directions, see Figure \ref{fig:normalPoints}), while Ganymede's and Europa's states are still poorly determined and could therefore affect Callisto's solution.}

\textcolor{black}{The results for the two above-described flybys are provided in Tables \ref{tab:considerParametersGanymedeFlyby} and \ref{tab:considerParametersCallistoFlyby}. They showed that including the other moons' states as consider parameters does not noticeably affect the formal errors obtained for the central moon's normal point. This confirms the validity of our simplified approach and allow us to keep excluding the non-central moons' states from the first step of the decoupled estimation. It must be noted that while this appendix only present only two flybys in detail, this analysis was conducted for all JUICE flybys, with identical conclusions.}

\begin{table*}[tbp]
	\caption{\textcolor{black}{Normal point determination for the 1st flyby at Ganymede, with and without including the other moons' states as consider parameters in the estimation. The formal errors in Ganymede's arc-wise state are provided in inertial coordinates.}}
	\label{tab:considerParametersGanymedeFlyby}
	\centering
	\begin{tabular}{l | c | c c | c}
		\hline
		\multicolumn{2}{l|}{\textbf{ Formal errors in }} &  \multicolumn{2}{c|}{\textbf{Other moons' states}} & \textbf{Relative difference }  \\ 
		\multicolumn{2}{l|}{\textbf{Ganymede's state}} & {Excluded} & {Consider parameters} & \textbf{[\%]}  \\ \hline
		& $x$ & 989 m  & 991 m & 0.18 \\
		Position& $y$ & 147 m  & 151 m & 2.8 \\
		& $z$ & 2.12 km & 2.21 km &  $\sim$ 0 \\ \hline
		& $v_x$ & 0.0645 m/s  & 0.0645 m/s& $\sim$ 0  \\ 
		Velocity& $v_y$ & 0.0104 m/s  & 0.0104 m/s & $\sim$ 0  \\
		& $v_z$ & 0.145 m/s  & 0.145 m/s & $\sim$ 0  \\
		\hline
	\end{tabular}
\end{table*}

\begin{table*}[tbp]
	\caption{\textcolor{black}{Normal point determination for the 12th flyby at Callisto, with and without including the other moons' states as consider parameters in the estimation. The formal errors in Callisto's arc-wise state are provided in inertial coordinates.}}
	\label{tab:considerParametersCallistoFlyby}
	\centering
	\begin{tabular}{l | c | c c | c}
		\hline
		\multicolumn{2}{l|}{\textbf{ Formal errors in }} &  \multicolumn{2}{c|}{\textbf{Other moons' states}} & \textbf{Relative difference }  \\ 
		\multicolumn{2}{l|}{\textbf{Callisto's state}} & {Excluded} & {Consider parameters} & \textbf{[\%]}  \\ \hline
		& $x$ & 1.69 m & 1.70 m & 1.4 \\
		Position& $y$ & 6.63 m & 6.63 m & $\sim$ 0 \\
		& $z$ & 126 m &  126 m & $\sim$ 0 \\ \hline
		& $v_x$ & $2.71\cdot10^{-5}$ m/s & $2.71\cdot10^{-5}$ m/s & $\sim$ 0 \\ 
		Velocity& $v_y$ & $4.59\cdot10^{-5}$ m/s & $4.59\cdot10^{-5}$ m/s & $\sim$ 0 \\
		& $v_z$ & $1.53\cdot10^{-3}$ m/s & $1.53\cdot10^{-3}$ m/s & $\sim$ 0 \\
		\hline
	\end{tabular}
\end{table*}

\setcounter{figure}{0}
\setcounter{table}{0}

\section{Deterministic simulation as verification for the covariance analysis} \label{appendix:verification}

While we limited ourselves to covariance analyses in our comparative study (see Section \ref{sec:scopeComparativeAnalysis}), we also performed a complete least-squares estimation as a verification of the coupled model, which we will refer to as a \textit{deterministic} simulation in the following. Our goal is to check whether the formal uncertainties discussed in our paper are consistent with the parameters values that would be estimated from non-ideal observations (\textit{i.e.} including biases and noises) using an iterative least-squares inversion. This is especially important given the high condition number encountered when inverting the normal equations  (see Sections \ref{sec:resultsFlybyPhase} and \ref{sec:resultsOrbitalPhase}).

\subsection{Scope of the verification}

\textcolor{black}{The deterministic simulations presented in this appendix remain within the scope of some major assumptions of the covariance analysis: in our simulated environment, this verification indeed still relies on \textit{perfect} dynamical and observational models. The additional simulations presented in the following thus do not hinder or replace the discussion in Section \ref{sec:modellingIssues} on dynamical modelling fidelity and its possible effect on the coupled and decoupled solutions. Deviations between the true errors and the statistical behaviour predicted by formal uncertainties mostly originate from observational and dynamical modelling inaccuracies. Therefore, this analysis is not meant as a validation and cannot investigate how representative \textit{formal} errors are of the \textit{true} errors that would be obtained from real data. It could be possible to purposely introduce perturbations in our dynamical models. However, simulating such errors such that their effects are representative of the limitations of our current models is not straightforward. It would require a dedicated, extensive analysis of our current dynamical and observational models, as discussed in Section \ref{sec:discussion}, and was therefore deemed beyond the scope of this comparative state estimation study.}

\textcolor{black}{Nonetheless, running a deterministic simulation still allows us to check that the \textit{formal} uncertainties are consistent with the \textit{true} errors resulting from the iterative least-square estimation, presuming that the assumptions of the covariance analysis hold. Such a verification ensures that our implementation of the estimation model is correct and  demonstrates that our formal uncertainties are reliable in a context of the covariance analysis.} 

\textcolor{black}{We only conducted such a deterministic simulation for the coupled model. It is indeed currently \textcolor{black}{less widely applied and less documented} than the decoupled estimation method, and such a verification thus becomes relevant.} 
\textcolor{black}{The coupled model being an extension of the classical formulations used by the decoupled estimation, the added value of performing a separate deterministic verification for the decoupled model appears limited. Additionally, implementing and running the entire least-squares estimation with the decoupled model would be more demanding, since it relies on two consecutive estimation steps (Figure \ref{fig:decoupledModel}) and in practice also requires multiple sequential estimations to determine all normal points, due to the chosen \textit{a priori} update strategy (see Section \ref{sec:aPrioriKnowledge}).}

\subsection{Approach and settings}

When performing an iterative weighted least-squares inversion, the variation in parameters values $\Delta\mathbf{q}_i$, at each iteration $i$, is given by \citep[\textit{e.g.}][]{montenbruckGill2000}
\begin{align}
\Delta\mathbf{q}_i &= \\
&\left( \mathbf{P}_{\mathbf{q}\mathbf{q},0}^{-1} + \mathbf{H}_i^{T}\mathbf{W}_i\mathbf{H}_i \right)^{-1}\left(\mathbf{H}_i^{T}\mathbf{W}_i\Delta\mathbf{z}_i+\mathbf{P}_{\mathbf{q}\mathbf{q},0}^{-1}\Delta\mathbf{q}_{0,i}\right), \nonumber
\end{align}
where $\Delta \mathbf{z}_i$ is the vector containing the observations residuals at iteration $i$, and $\Delta\mathbf{q}_{0,i}$ is the difference between the current parameters estimates and their \textit{a priori} values. The rest of the notation follows that of Equation \ref{eq:covariance}.

In our simulations, \textit{true} errors are directly defined as the difference between the \textit{true} parameters values (\textit{i.e.} values assumed in our simulated environment) and their estimated values. To keep our simulation as realistic as possible, the observations are modelled with the bias and noise levels given in Section \ref{sec:estimationSettings}: \textcolor{black}{noise levels of 20 cm, 15 $\mathrm{\mu m/s}$ and 0.5 nrad for range, Doppler and VLBI observations, respectively, with biases of 1 m for range measurements and 0.5 nrad for VLBI data}. \textcolor{black}{We also applied small initial perturbations to the \textit{a priori} values of the estimated parameters}, to mimic imperfect knowledge of the \textit{true} parameters values. The way the initial parameters' offsets were defined is detailed below:
\begin{itemize}
	\item moons' initial states: initial perturbation $\Delta \mathbf{q}_0$ calculated such that, once propagated over the duration of the JUICE mission, the difference between the perturbed  and unperturbed moons' trajectories would be of the order of magnitude of the current ephemerides' accuracy (see Section \ref{sec:estimationSettings}). \textcolor{black}{This led to initial offsets of 100 m in position and 5 mm/s in velocity.}
	\item JUICE's arc-wise states: offset of 100 m in position and 0.01 m/s in velocity (in all inertial directions);
	\item other parameters \textcolor{black}{(gravitational parameters, gravity coefficients, accelerometer calibration biases, range and VLBI observation biases)}: perturbation set to 5\% of their \textit{true} values.
\end{itemize}

Some differences should be noted with respect to the nominal estimation settings used for our covariance analyses (Section \ref{sec:estimationSettings}). Io's initial state was removed from the list of estimated parameters, due to the absence of Io's flyby, resulting in a lack of observations at this moon. The least-squares inversion could indeed not converge when trying to estimate its state alongside with those of the other Galilean moons. This already provides valuable insights about the challenges one will face when conducting real data analysis (as further discussed in Section \ref{sec:discussion}). It should be noted that this convergence issue is only caused by JUICE's imbalanced data set, and that many more difficulties are to be expected when dynamical and observational models will be confronted with real data (see discussion in Section \ref{sec:discussion}). \textcolor{black}{It is also important to highlight that this was only observed when perturbing the \textit{a priori} values of the estimated parameters. When no initial perturbation is applied, the least-squares estimation could converge while solving for Io's state along with the other moons'. All true errors were then found to be smaller than 3$\sigma$, $\sigma$ designating the corresponding formal errors.}

Furthermore, the elevation and occultation checks normally performed to verify the viability of an observation were switched off, thus assuming constant link during the tracking arcs. One specific tracking arc did not have enough viable observations for the estimation problem to converge otherwise. While this issue could have also been fixed with tighter \textit{a priori} constraints, we adopted a simpler approach. We must stress that removing the visibility requirements, thus adding a few observations, does not lessen the relevance of our verification analysis, since it focuses first and foremost on the consistency between true and formal uncertainties provided by the coupled model, and not on absolute error values.

We conducted this iterative least-squares inversion for the flyby phase case only. This was mostly motivated by the high computational load required by such deterministic runs. We do expect similar results and conclusions when including the orbital phase at Ganymede.  

\subsection{Results}

\begin{figure*}[tbp!]
	\centering
	\makebox[\textwidth][c]{\includegraphics[width=1.1\textwidth]{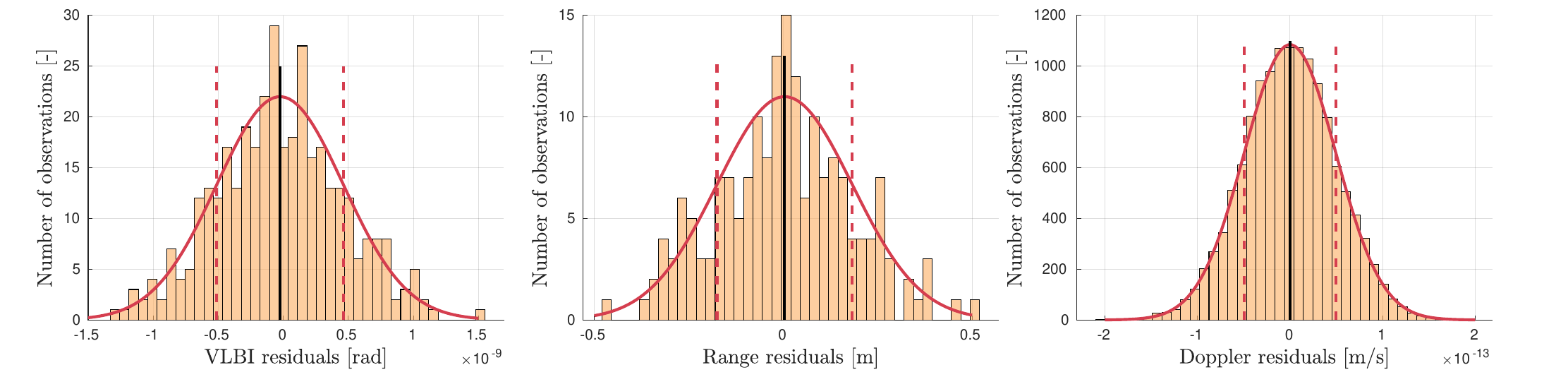}}
	\caption{Observations residuals after the weighted least-squares estimation reached convergence (from left to right: VLBI, range and Doppler residuals). The black vertical lines indicate the mean of the residuals for each observation type, while the dashed vertical lines represent the standard deviation around the mean value.}
	\label{fig:residuals}
\end{figure*}

\begin{figure}[tbp!]
	\centering
	\includegraphics[width=0.5\textwidth]{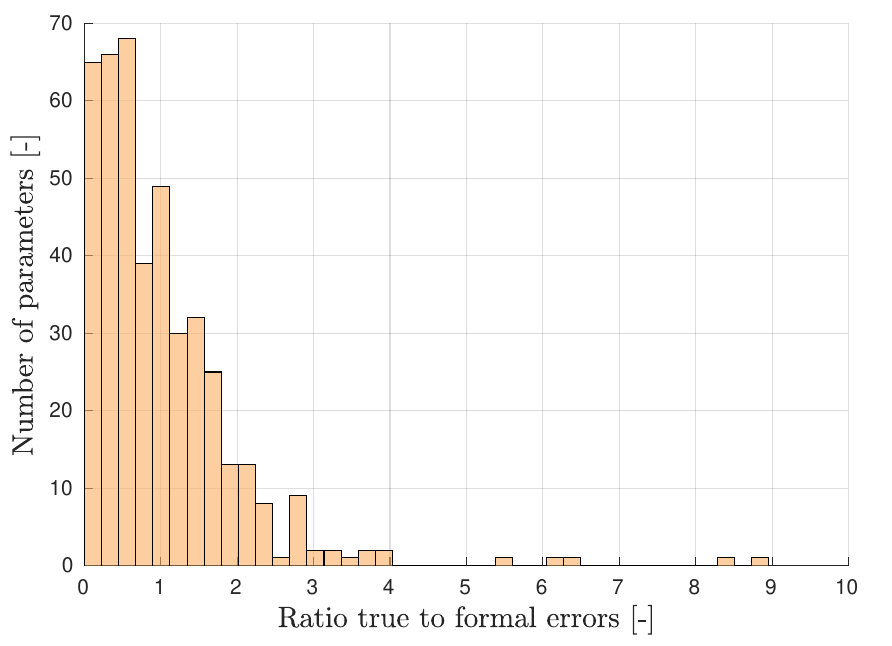}
	\caption{Distribution of the ratio between true and formal errors for all estimated parameters. A value smaller than 1 indicates that the true estimation error is lower than the associated formal uncertainty. 432 parameters were estimated in total.}
	\label{fig:ratioTrueToFormalErrors}
\end{figure}

After 5 iterations, the least-squares inversion reached convergence, and the final observations residuals are provided in Figure \ref{fig:residuals}. As expected, the residuals follow a Gaussian distribution with almost zero mean, and a standard deviation close to the observations noise level (Section \ref{sec:estimationSettings}). This can most clearly be observed for the Doppler residuals, due to the larger number of observations available compared to range and VLBI data. 

The histogram of the ratios between true and formal errors is displayed in Figure \ref{fig:ratioTrueToFormalErrors}, and all ratio values are between 0 and 9. Precisely quantifying which true to formal errors ratio should be expected is far from straightforward, as it is both data- and parameter-dependent. While true errors 2 to 3 times larger than formal errors can be typically expected for planetary ephemerides \citep[][for Saturn's ephemeris derived from VLBA observations]{jones2015,jones2020}, larger true to formal errors ratios have been found for dynamical parameters estimated from spacecraft tracking data (gravity fields, rotational parameters, \textit{etc.}) \citep[\textit{e.g.}][]{konopliv2011, mazarico2015}. These results were however all based on real observations, such that inaccurate dynamical or observations error models significantly contributed to the true/formal errors discrepancy. 

In our simulations, however, our models are still deemed perfect, and true errors should be comparable to formal uncertainties if the estimation is able to converge towards the correct parameters values. In other JUICE simulation studies, true errors were indeed found to be similar or even slightly lower than formal ones for Callisto's gravity field in \cite{diBenedetto2021}, although it is unclear if the parameters values were initially perturbed, and by how much. In \cite{lari2018b}, similar results were obtained, but true errors were averaged over 10 experiments with different initial parameters perturbations (smaller than those we applied).

In our simulation, for most parameters (97\% of the 432 estimated parameters), the true error is smaller than 3$\sigma$ (\textit{i.e.} true to formal errors ratio smaller than 3). A significant fraction (62\%) of the true errors are actually smaller than 1$\sigma$. These results confirm that, for the vast majority of estimated parameters, the coupled estimation model is able to properly converge towards the \textit{true} parameter value, and that errors can be expected to be in agreement with $3\sigma$ uncertainties. \textcolor{black}{All parameters exhibiting a true to formal errors ratio larger than 3 were actually found to be Europa-related properties (gravity field coefficients, initial position along the x-axis, \textit{etc.}). Due to the limited amount of observations collected at Europa (only two flybys, see Figure \ref{fig:flybysAltitudes}), these parameters are either highly correlated (second degree and order gravity field coefficients with Europa's state), or cannot be estimated beyond their \textit{a priori} constraints ($c_\mathbf{q}$ close to 1 for Europa's other gravity coefficients, implying a so-called \textit{biased} estimation). This might at least partially explain why unexpectedly large true to formal errors ratios were obtained for these parameters.} 

Overall, this deterministic simulation succeeded in verifying that the $3\sigma$ uncertainties provided by the coupled estimation are good indicators of the expectable estimation errors. This increased confidence in our covariance analyses. The effect of inaccuracies in dynamical or observational modelling was however not investigated (see discussion in Section \ref{sec:discussion}). Interestingly, and despite still relying on ideal models, this simulation nonetheless highlighted issues arising when trying to complete the least-squares estimation, due to the high condition number resulting from the lack of observations at Io and Europa.

\end{document}